\documentclass[printer]{aa} 
\usepackage{graphicx}
\usepackage{natbib}

\bibpunct{(}{)}{;}{a}{}{,} 
\begin{document}

   \title{A study of high velocity molecular outflows\\with an up-to-date sample
        \thanks{Table 1 is only available in electronic at the CDS via anonymous ftp to
            cdsarc.u-strasbg.fr (130.79.125.5) or via http://cdsweb.u-strasbg.fr/Abstract.html}
            }


   \author{Yuefang Wu\inst{1} \and Yue Wei\inst{1} \and Ming Zhao\inst{1} \and Yong Shi\inst{1} \and %
           Wentao Yu\inst{1} \and Shengli Qin\inst{1}\inst{3} \and Maohai Huang\inst{2}\inst{3}
          }

   \offprints{Yuefang Wu, \email{yfwu@bac.pku.edu.cn}}

   \institute{Astronomy Department, CAS-PKU Joint Beijing Astrophysics Center, %
              Peking University, Beijing 100871, China                          %
              \and Osservatorio Astronomico di Trieste,%
                   via Tiepolon. 11, Trieste 34131, Italy %
              \and National Astronomy Observatory of China, %
                   Beijing 100012, China}

   \date{Received ?, ?;  accepted ?, ?}

   \abstract{
            A statistical study of the properties of molecular outflows is performed based
            on an up-to-date sample. 391 outflows were identified in
            published articles or preprints before February 28, 2003.
            The parameters of  position, morphology, mass, energy,
            outflow dynamics
            and  central source luminosity are presented for each outflow source.
            Outflow lobe polarity is known for all the sources,  and 84\%
            are found to be bipolar. The sources are divided into low mass and
            high mass groups according
            to either the available bolometric luminosity
            of the central source or the outflow mass.
            The pace of discovery of outflows over the past seven
            years has increased much more rapidly than in previous
            periods. Surveys for outflows are still continuing.
            The number of high-mass outflows detected (139) has considerably
            increased, showing that they are commonly associated with
            massive as well as low mass stars.
            Energetic mass ejection may be a common aspect of the
            formation of high mass as well as low mass stars.
            Outflow masses are correlated strongly with bolometric luminosity of the
            center sources, which was obtained for the first time.
            There are also correlations between the central source luminosity and the
            parameters of mechanical
            luminosity and the thrust or force necessary to drive the outflow.
            The results show that flow mass, momentum and energy depend on the
            nature of the central source. Despite their similarity,
            there are differences between the high mass and low mass outflows.
            Low mass outflows are more collimated than high mass outflows.
            On average, the mass of high mass sources can be more than two orders
            of magnitude larger than those of low mass outflows. The relation
            between flow mass and dynamical time appears to differ for the two
            types of outflows.
            Low mass sources make up 90\% of outflows associated with HH
            objects while high mass outflows make up 61\% of the sources
            associated with H$_2$O masers.
            Sources with characteristics of collapse or infall comprise 12\% of the entire
            outflow sample.
            The spatial distribution of the outflow sources in the Galaxy is presented
            and the local occurrence rate is compared with the stellar birth rate.

            }

\authorrunning{Yuefang Wu et al.}
\titlerunning{A study of high velocity molecular outflows}
\maketitle


\section{Introduction}
    As one of the most exciting discoveries in astronomy
    in the past three decades,
    high velocity molecular outflows continue to attract
    great attention from researchers.
    Since their first discovery in 1976~\citep{Zuckermann_KK76, Kwan_S76,
    Zuckerman_P75},
    outflows have been detected at a high rate.
    67 had been discovered  by 1984 ~\citep{Lada85},
    144 by 1989~\citep{Fukui89},
    and 264 by 1996 \citep{Wu_HH96}.

    Outflows were thought to be the earliest observable signatures of star formation.
    Although evidence of mass loss for young stars had been detected previously
    in the optical spectrum \citep{Herbig60, Kuhi64},
    a better understanding of such mass loss in early
    evolutionary phases was only acquired after the discovery
    of molecular outflows in the material surrounding the
    embedded young stellar objects (YSOs).
    Molecular outflows show various characteristics
    such as radio and optical jets, atomic hydrogen stellar wind,
    molecular hydrogen jets and interstellar masers.
    The stars eject mass in energetic flows during formation~\citep{Lada85,
    Lizano_HR88, Natta_GP88, Lane89, Felli_PT92, Russell_BP92,
    Matzner_M99, Bally_JJ01, Reipurth_HM02}.
    In the 1990s, progress was made in observing and understanding
    the collapse of protostars \citep{Zhou92, Evans99, Myers_MT96}.
    According to theoretical models, collapses should occur before
    outflow~\citep{Shu_AL87}.
    However, nearly all the candidates of collapsing protostars are found in outflow sources.

    When advancement of observational techniques allowed the detection
    of high velocity molecular outflows by high J
    transitions,
    extremely high-velocity molecular outflows or highly collimated jet-like outflows
    were studied extensively
    \citep{Koo89, Bachiller_MP91, Choi_EJ93, Narayanan_W96}.

    More recently, considerable progress in outflow detection
    has been made in distant and complex high-mass star formation
    regions \citep{Beuther_SS02a, Zhang_HB01, Shepherd_C96a, Wu_YL99}.
    In the surveys of high-mass star formation regions conducted by
    \citet{Beuther_SS02a} and \citet{Zhang_HB01},
    the outflow detection rates reached 80\% and 90\%, respectively.
    Recent studies show that the sources in these surveys
    are either precursors of ultracompact (UC) HII regions,
    or are at very early evolutionary stages~\citep{Molinari_TR02, Beuther_SM02b}.

    The outflow parameters and their correlations have been
    investigated \citep{Rodriguez_CH82, Bally_L83, Cabrit_B92, Shepherd_C96b,
    Ridge_M01}.
    Mechanisms of outflow collimation and the forces driving the outflows
    have been proposed as well.
    However, the correlation between morphology, physical parameters
    and the central source conditions are still not well understood.
    For example, it is not clear if the collimation is
    correlated with the driving force.
    The correlation among
    bolometric luminosity
    of the center source, the required force and mechanical luminosity
    was investigated by \citet{Bally_L83} and \citet{Rodriguez_CH82}.
    As the correlation was discovered to be weak, a larger sample is required.

     To further understand the mechanisms of high velocity outflows,
     we conducted an analysis
     based on an up-to-date sample. Our calculations and conclusions are presented
     in this paper.

     The used data in this paper are heterogeneous.
     The outflows were detected at different stages of the young stellar
     objects: high mass sources from pre-UCHII region to UCHII region and
     low mass sources from Class 0 to FU Orionis.
     The observational sensitivity and the calculation method also
     varied from source to source.
     However, we believe that our sample is sufficiently large
     to provide the overall physical characteristics of outflows.
     Sect. \ref{catalogue} presents the sample, the catalogue and a brief analysis.
     The morphology analysis in Sect. \ref{morph}.
     Sect. \ref{phy} discusses the physical parameters.
     The associated objects or phenomena of outflows are discussed
     in Sect.~\ref{associated}.
     Sect.\ref{distri} describes the spatial distribution of outflows.
     We give a summary in Sect. \ref{sum}.

\section{The catalogue\label{catalogue}}

    \subsection{The sample}
        We compiled outflow sources mapped mainly
        in emission lines of low transitions
        ($J=1-0$ and $J=2-1$) of CO showing evidence of large scale red and blue lobes.
        The catalogue of 1996 \citep{Wu_HH96} included 28 sources without CO
        maps (with superscript a in the source name).
        We exclude those sources that have not been mapped so far.
        If an outflow was detected with CO $J=3-2$ or higher transition lines,
        it is noted in the table accordingly.
        Some authors detected molecular outflows with other molecular species,
        such as SiO, CS, HCO$^+$
        \citep[e.g.][]{Megeath_T99, Wolf-Chase_BW98, Garay_KB98, Hofner_WH01}.
        These species are not included in this catalogue.

        The sources presented in this new catalogue were identified
        as high velocity molecular outflows
        in published findings or in  preprints published before February 28, 2003.
        There are 397 sources; 391 show single outflows,
        6 were observed as single outflows but were shown later to be of multiple sources.
        These six were presented in the catalogue, but are not included in our analysis.

    \subsection{The tables}
        Table 1a and 1b list the high velocity molecular outflows
        and their parameters.
        Some of the outflows have been investigated by multiple authors,
            and have been mapped more than once.
        For these sources several entries are listed.
        Each entry corresponds to the references listed in the last column of Table 1b.
        There are 24 columns for each entry.
        Column 1-13 are in Table 1a, and the remaining columns are in Table 1b.
        The sources are presented in order of right ascension.

        Table 1a contains the basic parameters of the outflows:

        \begin{itemize}
            \item Column 1 (No.) is the sequence number of the outflow.
            \item Column 2 (Name) the source name and the alternative
name if any.
            \item Column 3 ($\alpha$) and 4 ($\delta$) are the right ascension and declination (1950.0).
            \item Column 5 ($l$) and 6 ($b$) are the galactic longitude and latitude.
            \item Column 7 ($D$) and 8 ($Z$) are the distance from the sun and altitude from the Galactic plane.
            \item Column 9 ($\Delta  V$) lists the full line-width at 0.1-0.2K above the baseline.
                 For extremely high velocity sources, an additional line-width
measured at tens of milli-kelvins is provided.
            \item Column 10 (Po.) shows the polarity of the outflow.
                Here, `Bi' represents bipolar outflow;
                `MB' and `MR' indicate blue and red monopolar outflows,
                `Multi' means multiple outflows and `Iso' isotropic outflows.
            \item Column 11 ($R_{\rm max}$) and 12 ($R_{\rm coll}$) present the size and the collimation factor, respectively.
                To calculate the collimation factor, we first consider an outflow as an ellipsoid.
                Then we measure the area and maximum angular extent (the major radius ) of the outflow,
                and calculate the minor radius. The ratio of the major and minor radii is the collimation factor \citep{Bally_L83}.
                The value underlined in column 12 is quoted from references where the same method was employed.
            \item Column 13 ($L_{\rm bol}$) shows the bolometric luminosity of the center source
                which is  the deriving source or its candidate,
                indentified by infrared photometry or by IRAS data.
        \end{itemize}

        The physical parameters are listed in Table 1b (column 14-24),
        mainly derived from lower level CO lines (CO $J=1-0$, $2-1$).
        If the parameters were calculated with CO $J=3-2$ or higher levels,
        we note them with `c'. For extremely high velocity outflows we only list the total velocity range.

        \begin{itemize}
            \item  Columns 14 and 15 contain source numbers and their names corresponding to
                   Table 1a column 1 and 2.
            \item Column 16 ($M$) is the outflow mass calculated by integrating CO emission
                  from the entire area of the wings \citep{Goldsmith_SH84, Snell_SS84, Lada85}.
                  The mass obtained under optically thick conditions are noted  `h'.
                  The rest are calculated in the optically thin regime.
            \item The momentum ($P$), kinetic energy ($E$) and dynamic time ($t$) of the outflows are
                  listed in column 17, 18 and 19, respectively \citep{Lada85}.
            \item The mechanical luminosity ($L_{\rm m}$) of the outflow and force ($F$)
                  derived from outflow physical parameters \citep{Bally_L83, Goldsmith_SH84, Snell_SS84}
                  are shown in column 20 and 21, respectively.
            \item Column 22 ($\dot M$) indicates the mass loss rate of the center stellar source,
                  which was calculated with the outflow parameters and
                  terminal velocity of the stellar wind
                  \citep{Bally_L83, Goldsmith_SH84, Snell_SS84, Levreault88}
                  and the corresponding reference of the outflow.
            \item Column 23 (IIH2WI) presents the phenomena associated with the outflows.
                  `II' stands for UCHII regions, `H' for optical jet or HH objects,
                  `2' for H$_2$ jet, `W' for water maser,
                  and `I' for infall or collapse candidates.
            \item Column 24 (Ref.) presents the references.
        \end{itemize}

    \subsection{Classification}

        The bolometric luminosity of the center source of the catalogued
        outflows ranges from one tenth  to more than
        $10^6$ times solar luminosity ($\rm L_{\odot}$).
        We adopt $L_{\rm bol}=10^3~{\rm L_{\odot}}$ as a criterion
        to distinguish between high-mass sources and low-mass
        sources.
        Young high-mass stellar sources are brighter than $10^3~\rm L_{\odot}$
        at minimum.
        On the H-R diagram, $L_{\rm bol}$ of low-mass and intermediate mass
        sources is mostly below $10^3~\rm L_{\odot}$
        \citep[ and references therein]{Levreault85}.
        The criterion of $10^3~\rm L_{\odot}$  was similarly employed
        in the comparative study of
        different mass molecular outflows or cores
        \citep{Shepherd_C96b, Fukui89, Wu_WW01}.

       Many sources have more than one set of parameters derived
       from different observations, therefore we classify them
       according to the following considerations.
       First we only classified sources whose multiple measurements
       of masses have consistent $L_{\rm bol}$ values either
       larger or smaller than 1000~$\rm L_\odot$.
       In the classified sources,
       292 sources have available $L_{\rm bol}$ measurements,
       of which 120 (41 \%) have $L_{\rm bol}>10^3~{\rm L_{\odot}}$,
       and 172 (59 \%) have $L_{\rm bol}\leq 10^3~{\rm L_{\odot}}$.
       The rest of the sources have no available $L_{\rm bol}$,
       but some have available outflow gas masses.
       Of the 113 sources with $L_{\rm bol}>10^3~{\rm L_{\odot}}$
       and with available masses,
       94 (82\%) have $M>3~{\rm M_{\odot}}$;
       for 138 sources with $L_{\rm bol}\leq 10^3~{\rm L_{\odot}}$
          and with available masses,
       111 (80\%) have $M\leq 3~{\rm M_{\odot}}$.
        Sources whose masses observed by different authors have values both
        larger or smaller than $3~{\rm M_{\odot}}$ were not taken
        into account in the above calculation.
        Thus, the two percentages cited above are lower limits.
        Further statistics show that the outflow mass is correlated with $L_{\rm bol}$ (see Sect.
         \ref{mass-lbol}).
         For sources without $L_{\rm bol}$ but with mass available,
            we use $3~\rm M_{\odot}$ as a criterion to classify them.
         Among these sources there are 19  with $M> 3~{\rm M_{\odot}}$ and 51
         with $M\leq 3~{\rm M_{\odot}}$.
        We refer to the 139 sources with $L_{\rm bol}>10^3~{\rm L_{\odot}}$ (120) or $M>3~{\rm M_{\odot}}$ (19)
            as the high mass group and the 223 sources with $L_{\rm bol}\leq 10^3~{\rm L_{\odot}}$ (172)
            or $M\leq 3~{\rm M_{\odot}}$ (51) as the low mass group in the following text.
        Currently we have  not found the mass or bolometric luminosity  for 29 sources.
        These sources need further examination.

    \subsection{Development of outflow detections}

        New outflows have been detected and  surveyed at an accelerated rate in the past
        two decades.
        There are now 397 identified
        sources (including 6 unresolved outflows).
        The number of sources is  seven times greater than that
        catalogued in 1985
       (which had at most 55 mapped)~\citep{Lada85},
        244 more than that of 1989~\citep{Fukui89},
        and 159 more than that of 1996~\citep{Wu_HH96}.

\begin{figure}
    \centering
    \resizebox{\hsize}{!}{\includegraphics[width=6cm]{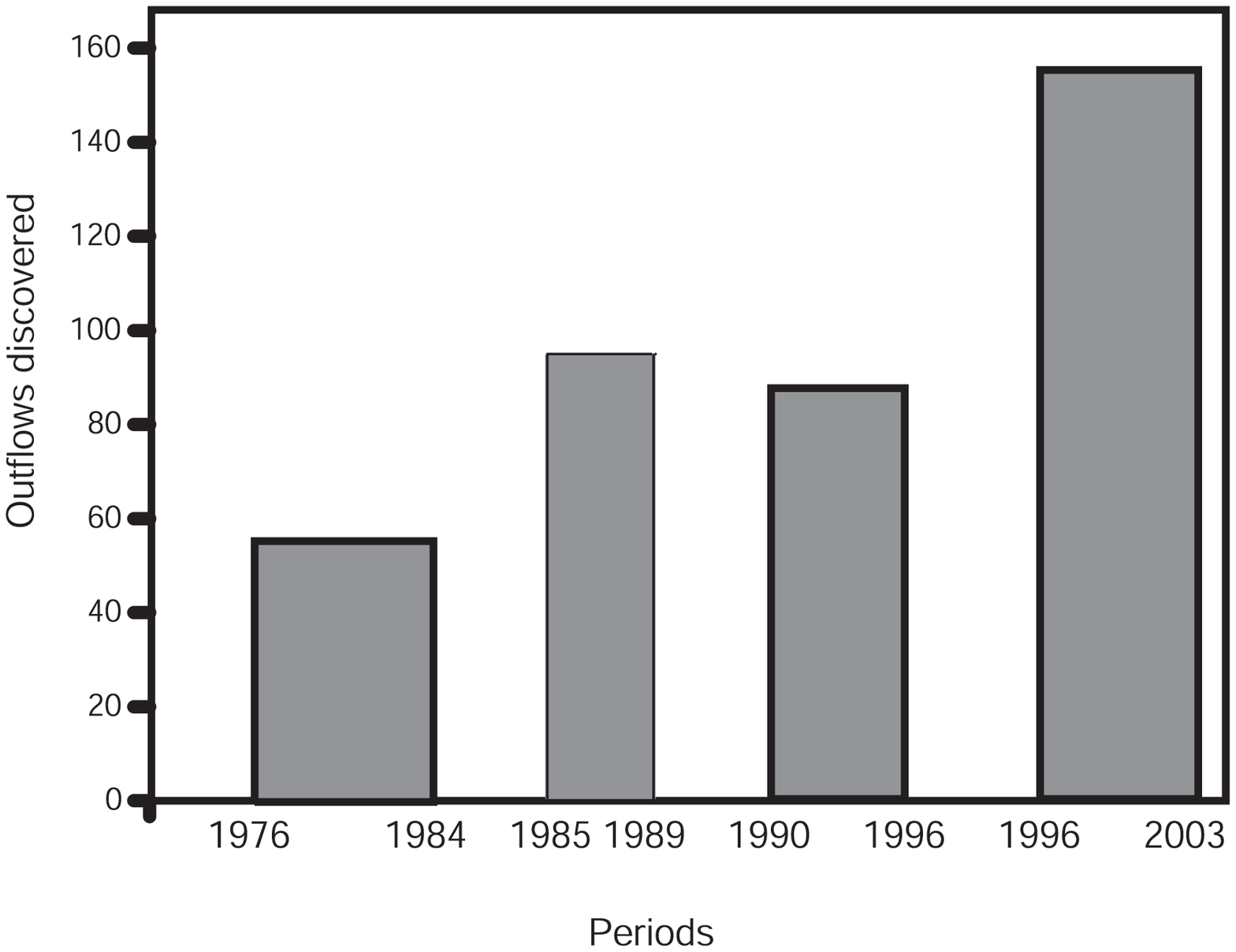}}
    a
    \resizebox{\hsize}{!}{\includegraphics[width=6cm]{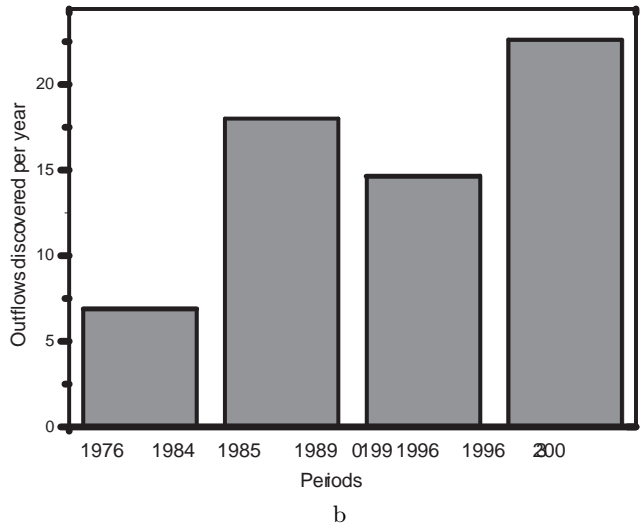}}
    b

    \caption{Frequency distributions of detected outflow numbers with time.
        a). Histogram of the mapped outflow numbers in the four periods; the
        widths of the shadow zones are determined by how long the corresponding
        period lasting;
        b). Histogram of average numbers per year in the four periods. }

    \label{periods}
\end{figure}

        Fig.~\ref{periods}a is the diagram of the number
        of outflows mapped in four periods, which is defined with
        the available catalogues of outflows.
        Period I is from 1976 to 1984 with 55 detected outflows~\citep{Lada85};
        Period II is from 1985 to 1989, with 95 outflows~\citep{Fukui89};
        Period III is from 1990 to 1995, with 88 outflows~\citep{Wu_HH96};
        Period IV is from 1996 to 2003 with 159 outflows (this work).
        The number of detected outflows has increased, especially in period IV
        (see Fig.~\ref{periods}a).
        Fig.~\ref{periods}b is the average number
        of mapped outflows in these four periods.
        The average number of outflows per year are 6.9, 19, 14.6
        and 22.7 for the four periods
        respectively.
        The trend of increasing number of detected outflows
        is evident,
        showing the development of outflow detection in recent years.

        In period II mainly low mass outflows were added \citep{Fukui89},
        while in period IV the number of high mass sources increased rapidly.
        Among the sources detected up to 1995, high mass sources occupy only 31\%
        (65 of 210 outflow sources in the high mass group with the same classification).
        Now there are 139 in the high mass group among 362 sources
        with available $L_{\rm bol}$ or $M$. The ratio is 38\%.
        The results show that high velocity molecular outflows
        are common in high mass star formation regions
        considering the fewer samples of high mass stars.
        It may suggest that high mass stars still form
        through an accretion-outflow dynamic process,
        like the formation process of low mass stars.

            Is the percentage of high mass here beyond that expected from the
            IMF? Here 38\% is the ratio of sources in the `high mass group'
            to the total number of sources in both the high and low
            mass groups. Part of the `high mass group' are
            actually intermediate mass sources. Considering the
            sources with $L_{\rm bol}>10^4 L_\odot$, the
            percentage drops to 23\% (83/362). The
            percentage of sources associated with UCHII regions (23\% (90/391))is
            also lower than 38\%.
            This percentage still exceeds what is expected from the
     IMF. But observed IMFs are derived from counting main-sequence
     stars in solar neighborhood~\citep{Kroupa_TG90, Salpeter55},
      According to Fig.~\ref{xyz}, massive outflows are easier to
      detect at increasing distance. Outside the 2 kpc circle,
      most outflows belong to the high mass group, while within 1
      kpc, there are only 10\% (16/167) high mass outflows.

        A notable trend in outflow research is that
        interferometers are being applied to measure the outflow regions.
        Massive outflows have been observed with higher spatial
        resolution and sensitivity in regions
        such as IRAS 05358+3543, G196.16 and NGC 7129
            \citep{Beuther_SG02c, Shepherd_WS98, Feunte_NM01}.

\section{Morphology and collimation\label{morph}}

     The polarity was known for all the outflows.
     327 (84\%) sources show bipolar structure;
     50 (13\%) sources are mono-polar, of which 28 are red polar.
     There are 12 (3\%) multi-polar sources;
     if a source was detected as bipolar first
     but was later found to have multiple polarities
     we count it as multi-polar.
     Two sources are still isotropic (No. 231 M8E and No. 391 MWC 1080).
     Compared with the sources cataloged before \citep{Wu_HH96, Lada85},
     the number of multi-polar sources has increased
     (there is only one in the previous periods)
     and the number of isotropic sources decreased (4 in Period I, 3 in Period III),
     which can be attributed to the enhanced spatial resolution of
     the detection equipment.

    The collimation factor is used to investigate the morphology of outflows.
    Different methods were employed to
    analyze collimation properties of outflows \citep{Cabrit_B92, Levreault85}.
    In this paper the collimation factor was obtained according to~\citet{Bally_L83} .
    Here we analyze the relationship between the collimation factor,
    the bolometric luminosity of the central source and the angle
    sizes.

    As shown in Table 1a, there are 213 sources
    with available collimation factors.
    The average value of the collimation factor
    for all the sources is
    2.45, with a standard deviation of 1.74,
    which is almost the same as that catalogued
    in period I and III~\citep{Lada85,Wu_H98}.
    The average collimation values for high mass
    and low mass group members are 2.05 and 2.81,
    with a standard deviation 0.96 and 2.16 respectively.
    The above values are similar to those in period III,
    1.99 and 2.54 for these two groups
    according to the same statistical method~\citep{Wu_H98}.
    These results show that the low mass outflows are better
    collimated than the high mass ones.
    Although the overall instrumental angular resolution
    in period IV has increased,

    the observed collimation has not improved much
    as a whole.
    This is likely due to the increased number of massive outflows.

\begin{figure}
    \centering
    \resizebox{\hsize}{!}{\includegraphics[]{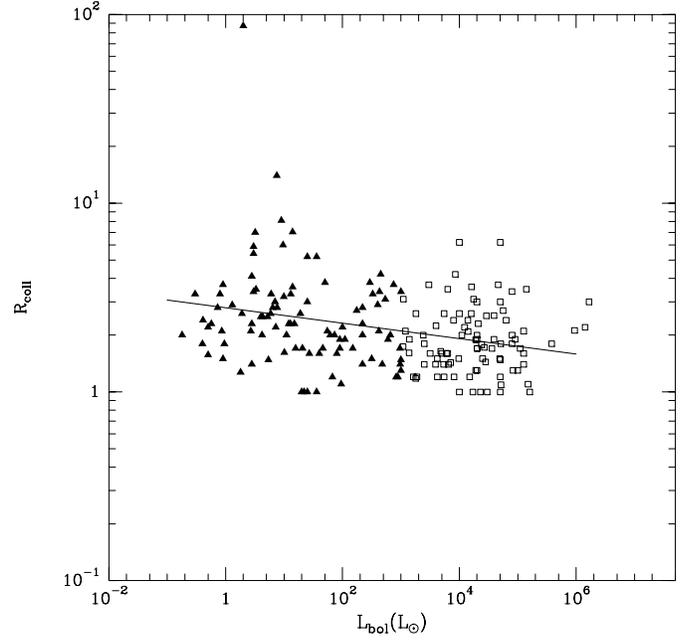}}

        \caption{Collimation factors versus bolometric luminosity of the
            associated infrared source ($L_{\rm bol}$).
            The filled triangles represent the sources from the low mass group,
            while the open boxes indicate the sources from the high mass group.
            The solid line is the least square linear fit line.}
     \label{Rcoll}

\end{figure}

\begin{figure}
    \centering
    \resizebox{\hsize}{!}{\includegraphics[]{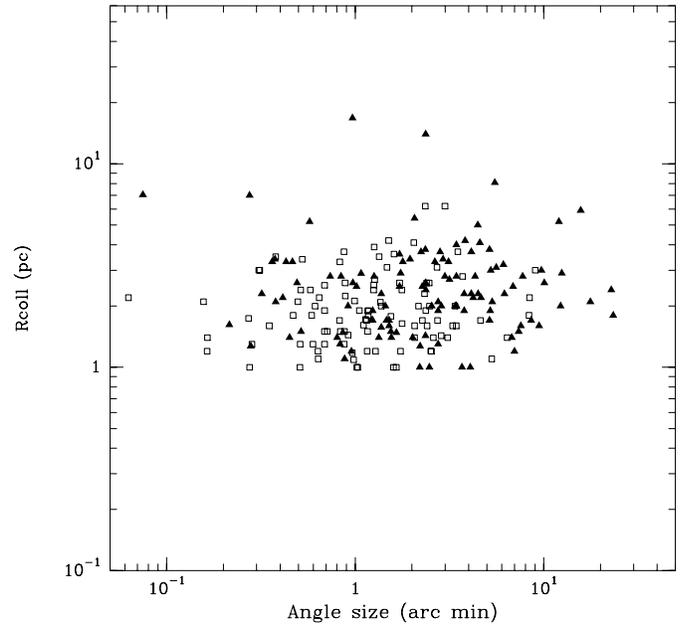}}

        \caption{Collimation factors versus angular sizes.
            The symbols are the same as for Fig.~\ref{Rcoll}.}
     \label{angular}
\end{figure}

\begin{figure}
    \centering
    \resizebox{\hsize}{!}{\includegraphics[]{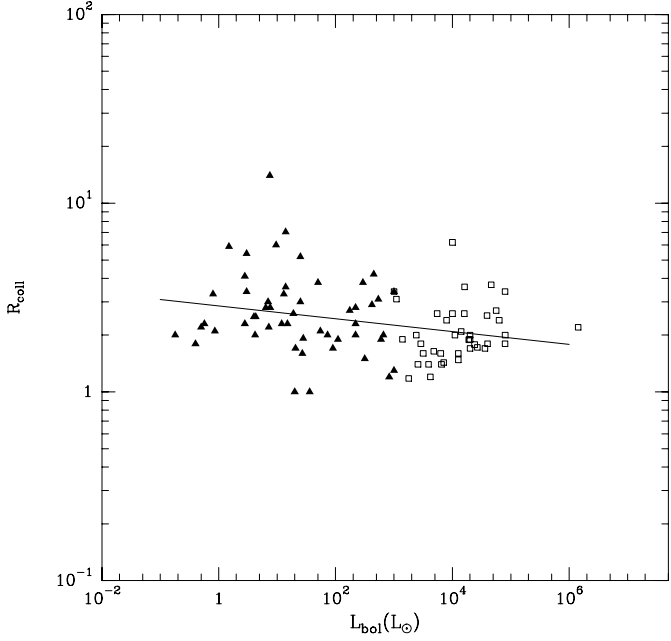}}

        \caption{Collimation factors versus bolometric luminosity ($L_{\rm bol}$)
            for sources with angular sizes at least five times the beam size
            The symbols are the same as for Fig.~\ref{Rcoll}.}
     \label{Rcoll5}

\end{figure}

    Figure~\ref{Rcoll} is the plot of $R_{\rm coll}$ vs. $L_{\rm bol}$.
    Solid triangles indicate low mass sources and squares indicate high mass ones.
    The solid line is a linear fit of the plots:
        $\log(R_{\rm coll})=(0.44\pm 0.03)+(-0.04\pm 0.01)\log(L_{\rm bol})$;
        the correlation coefficient is poor, $R=-0.29$.
    There is a slight difference between high mass and low mass outflow collimations.
    To analyze the effect of angle size on the collimation factor,
        we plot $R_{\rm coll}$ vs. the angular size in Fig.~\ref{angular}.
    It shows that there is no obvious correlation between
    collimation factor and angular size.
    To further investigate the relation between $R_{\rm coll}$ and $L_{\rm bol}$,
    an analysis was made of all sources whose angle size
    is five times larger than the beam size.
    Figure~\ref{Rcoll5} plots the $R_{\rm coll}$ vs.
    the $L_{\rm bol}$ for these sources.
    The solid line is the linear least-square fit:
    $\log(R_{\rm coll})=(0.46\pm 0.04)+(-0.03\pm 0.01)\log(L_{\rm bol})$;
    the correlation coefficiency is also poor, $r=-0.28$.
     The average values of $R_{\rm coll}$
     for the sources with large angle sizes are 2.16
     with a standard deviation 0.91 for high mass sources
     and 3.00 with a standard deviation 2.02
     for low mass sources.
     Both are better than those before the effect of beam size was removed
     ($2.05\pm 0.96$ and $2.81\pm 2.16$ for the two group sources).
     It shows again that the collimation degree tends to be lower
     with increasing bolometric luminosity of the centre source.
     However, the weak correlation between $R_{\rm coll}$ and $L_{\rm bol}$
     remains for sources with large angle sizes,
       as Fig.~\ref{Rcoll} shows.
     This may be related to the driving processes of the outflows.
     Outflows driven by wide-angle wind may be less collimated than those driven by jets.
     Now it is believed that low mass outflows are driven by bow shocks
     \citep{Lee_MR00,Chernin_M95,Raga_C93,De_Young86}
     while wide-angle wind may drive massive outflows \citep{Molinari_TR02}.
     Low angular resolution also made it difficult  to see the fine
     structure of multi-flows.
     For example, IRAS 05358+3543, detected as one outflow, was
     separated into at least three outflows by the Plateau de Bure
     interferometer~\citep{Beuther_SG02c, Beuther_SS02a},
     including one highly collimated outflow.
     However, this is not to say that instruments
     determine the values of collimation.
     Of the sources depicted in Fig.~\ref{angular},
     some small angle size outflows still have good collimation.

    The correlation between $R_{\rm coll}$ and $L_{\rm bol}$
    is not as obvious as that between outflow mass
    and bolometric luminosity (see next section).
    Both Fig.~\ref{Rcoll} and~\ref{Rcoll5} show
    that for collimation there is no definite
    dependance on the bolometric luminosity of the driving sources.
    Besides the sensitivity and resolutions of the observation
    equipment,
    several  effects may affect the measurement of collimation degree
    of the outflows.
    For example, changing the projection along the
    line of sight can change the value
    of the apparent  collimation factor for the same bipolar outflow to be smaller
    or larger than the intrinsic degree of collimation~\citep{Wu_H98}.
    Several factors may influence the intrinsic degree of collimation.
    Physical conditions of the environment will affect flow
    morphology.
    Recently a numerical simulation showed that
    environmental structure may determine the collimation
     degree   of observed outflows of massive YSOs \citep{Bonnell_B02}.
    Evidence was also found for the existence
        of an aligned magnetic field in the
        outflow regions \citep{Greaves_HW01, Houde_TP01}.
    Lastly, collimation may also be related to
    the evolution stage of the central object,
        and to the acceleration or deceleration
        phase of the flow itself \citep{Cabrit_B86, Cabrit_B90, Shu_AL87}.
     The outflows in Class 0 objects with a cold
     black body spectral energy distribution (SED)
        are more collimated than those in Class I objects
        whose SED is wider than that of a black body
        and whose positive indices show a
        more evolved status \citep{Bachiller96, Bontemps_AT96, Gueth_G99}.

\section{Physical parameters\label{phy}}

    Although the physical parameters of outflows have wide ranges,
    we believe that statistics derived from a large sample
        should still reveal general characters and correlations of
        the parameters of the outflows and indicate their driving mechanisms.
    $R_{\rm max}$ ranges from 0.01 to 3.98 pc;
    mass from $10^{-3}$ to $10^3 \,\rm M_{\odot}$;
    momentum from $10^{-3}$ to $10^4 \rm \,M_{\odot} km/s$;
    energy: $10^{38}$  to $10^{48} \,\rm ergs$;
    dynamic times $t$ from $4\times 10^2$ to $6\times 10^5 \,\rm yr$;
    mechanical luminosity $L_{\rm m}$ from $10^{-5}$ to $10^3\,{\rm L_{\odot}}$;
    force $F$ from $10^{-7}$ to $10^2\,{\rm M_{\odot}\,km/s\,yr}$;
    mass loss rate $\dot{M}$ from $10^{-9}$ to $10^{-3}{\rm M_{\odot}/yr}$.
    (Notice that the limits of different parameters may not come from
    the same source.)
    The property value ranges are wider than those of the sources measured
        in a single survey \citep{Bally_L83, Levreault85,
        Fukui89, Ridge_M01, Zhang_HB01, Beuther_SS02a};
        these results from the  fact that multiple surveys
        for different mass objects are included in our sample.
    Our sample also includes the most massive and the smallest outflows (see Sect.~\ref{LL}).
    However  sources with parameters near the limit are rare.

    We estimate the maximum possible errors that different considerations
    in deriving physical parameters may bring.
    The outflow emission at line wings is often accepted as
    being optically thin. Nevertheless, some authors deal with it using
    a derived or assumed optical depth, which may bring
    a factor of about 5 to the final value of outflow mass, momentum,
    kinetic energy and so on~\citep{Snell_SS84, Goldsmith_SH84, Shepherd_C96b, Garden_HH91, Yu_BB99}.
    Different authors used different $\rm [CO]/[H_2]$ abundance
    ratios.
    Usually this will affect the derived masses by a factor
    of less than 2~\citep{Ridge_M01}. Occasionally it could be
    larger, from $2.5\times 10^{-5}$~\citep{Rodriguez_CH82} to
    $10^{-4}$~\citep{Garden_HH91}.
    We accept a factor of 4 as the
    abundance error.
    Some authors use 1.36 as the mean atomic weight of the mixture of
    hydrogen and helium (see ~\citet{Garden_HH91}),
    while others only consider pure hydrogen molecular gas (see~\citet{Snell_SS84}).
    Therefore an uncertainty  of 0.36 is introduced.
    Projection effects affect momentum $P$ and energy $E_{\rm k}$
    by factors of 2 and 3 respectively~\citep{Goldsmith_SH84}.
    Our experience shows that different velocity ranges ($\Delta V$) of
    the determined outflow emission  add $20\%$ uncertainty to
    the results.
    The excitation temperature, $T_{\rm ex}$, is determined by the
    relative line intensities of at least two different
    transitions~\citep{Garden_HH91}.
    It is usually assumed or defined as the brightness temperature of the CO line peak.
    $T_{\rm ex}$ ranging from 10 to 15 K for low mass sources
    \citep{Goldsmith_SH84, Snell_LP80}, and from 30 to 50 K
     for high mass sources~\citep{Shepherd_C96b,
    Shepherd_C96a, Beuther_SS02a, Wu_04},
    causes a maximum uncertainty of about $60\%$.
    To investigate how different telescope beam sizes affect the
    outflow physical parameters, we pick out the sources whose
    angular diameters are larger than 5 beam sizes of the telescope
    used. 113 of these sources have available $L_{\rm bol}$ and
    $M$.  the linear fit of $\log L_{\rm bol}$ and $\log M$ is:
    $\log M =(0.92\pm 0.14) + (0.51\pm 0.04)\log L_{\rm bol}$,
    $r = 0.73$.
    Comparing the $L_{\rm bol}-M$ relation here to that of all the
    235 sources with available $L_{\rm bol}$ and $M$, which we
    present in Sect.~\ref{mass-lbol}, we find that:
    The slope of $> 5 $ beam sizes sample (0.51) is a little smaller
    than that of all (0.56), mainly because the majority of
    outflows more massive than $10^2 {\rm M_{\odot}}$ disappeared
    in the $>5$ beam sizes sample due to their greater distances.
    The correlation coefficients of the $>5$ beam sizes sample
    (0.73) is also a little worse ($r=0.78$ for all), because the
    $>5$ beam sizes sample is  less than  half of the whole set.
    However, the above two deviations are not significant. Thus we
    are able to conclude that the telescope beam size affects
    little  the outflow physical parameters.
    We also examine the differences which different CO $J$ transition
    may cause. The mass derived from CO $J=1-0$ and $2-1$ are expected
    to be within a factor of 2 to 3 of the actual values ~\citep{Margulis_L85},
    while this difference mainly comes from the optical depth difference which is
    already accented in our uncertainties.
    As a result, the outflow mass, $M$, has an uncertainty
    factor of 6.4;
    the outflow force, $F$, 6.7;
     the outflow luminosity,
    $L_{\rm m}$, 7.1.
    The bolometric luminosity, $L_{\rm bol}$,
    listed in column 13, primarily derived
    from far-infrared observations such as IRAS, should be fairly
    accurate~\citep{Lada85}. Even if there were more than one YSO
    within the IRAS beam, the emission of the most massive YSO will
    be the main contribution to the total FIR flux according to
    the mass-luminosity relation~\citep{Casoli_CD86}.
    \citet{Cabrit_B92} listed $L_{\rm bol}$ of 16 outflow sources.
    We compared these $L_{\rm bol}$ values to those listed in our
    Table 1a,
    and found that the uncertainty is within 2$\sim$3 except for
    Orion-A ((1$\sim$10)$\times 10^4$ and 2.1$\times 10^5$)and S140 ((0.5$\sim$9.0)$\times 10^4$ and 5$\times
    10^3$).
    For most cases, the variation of luminosities provided by different authors
    in our sample is about  2 to 3, except for a few high mass sources.
    The errors estimated
    above are much less than the scatter of the data themselves.

    \subsection{Relation between the outflow mass and the bolometric luminosity of the center sources\label{mass-lbol}}

        A plot of outflow mass relative to the bolometric
        luminosity of driving sources shows that the two
        parameters are
         physically related.
        The mass of the outflow is a fundamental parameter.
        Its relationship to the luminosity of the central source is
        essential to investigate the nature of the outflows.

\begin{figure}
    \centering
    \resizebox{\hsize}{!}{\includegraphics[]{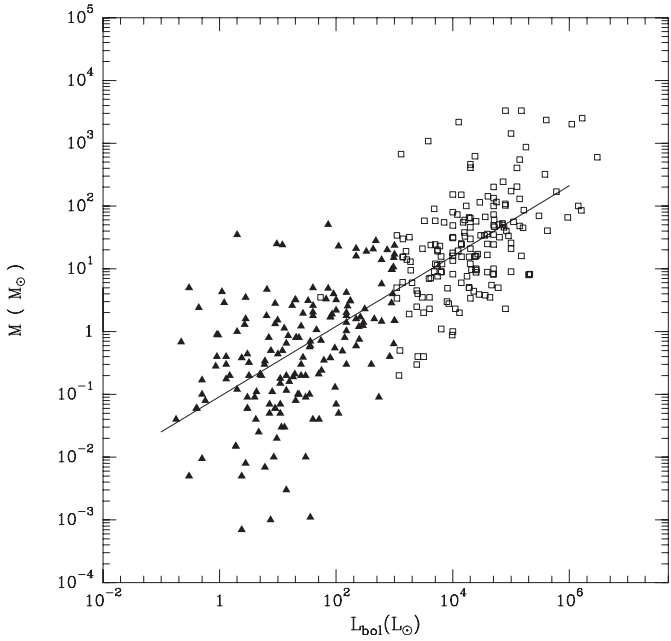}}

        \caption{The outflow mass versus the bolometric luminosity ($L_{\rm bol}$).
            The symbols are the same as in Fig.~\ref{Rcoll}.
            The solid line is the least square linear fit.}
     \label{LbolM}

\end{figure}

        Figure~\ref{LbolM} plots the mass of outflows $M$ vs.
            the bolometric luminosities of center sources $L_{\rm bol}$.
It shows that the flow mass
        increases with the luminosity as a power law.
        The outflows with higher luminosity values tend to have larger flow masses
         and vice versa.
        The least square linear fit in log-log space shown in the figure is
            $\log M= (-1.04\pm 0.08)+(0.56\pm 0.02)\log L_{\rm bol}$
            with the correlation coefficient $r=0.78$.
        The strong correlation between outflow mass and luminosity
        is an interesting result 
        which is the first demostration of such a clear
        correlation.   
        It presents essential clues to the origin
        and properties of outflows.
        The correlation suggests that the outflow mass is
        linearly correlated with the bolometric luminosity of the center source.
        The outflow however could not be driven directly by radiation pressure
        if the photons emitted from the center objects are scattered once
        before they escape \citep[see also Sect.~\ref{LL}]{Bally_L83}.
        This is a classic example in which two parameters
        are correlated because individually     
        each one is correlated to a third underlying parameter.
        In our case, it could be that the bolometric luminosity is
        correlated to the accretion rate
        ~\citep{Kenyon_HS90, Ohashi_KI91, Mundy_WW92, Greene_WA94},
        which in  turn is
        correlated to the mass loss rate in the outflow,
        \citep{Tomisaka_98, Shu_NO94, Contopoulos_S01, Fernandez_C01}
        and the
        mass loss rate in the outflow sources is what determines
        the mass swept up in the outflow lobes
        \citep{Snell_LP80, Ho_MR82, Bally_L83, Goldsmith_SH84}.  
        So although the radiation energy that the source provided
        is not able to drive such an outflow mass,
        they are still correlated.   

    \subsection{Relation between the outflow mechanical luminosity, required force
         and the bolometric luminosity of the center sources\label{LL}}

        To investigate the relationship between outflows and their central sources,
        we further plot the mechanical luminosity ($L_{\rm m}$)
        of outflows vs. bolometric luminosity ($L_{\rm bol}$)
        of the center stellar objects, shown in Fig.~\ref{Lbollm}.
        The dashed line shows the equation $L_{\rm m} = L_{\rm bol}$.
        The plot shows that for all sources, $L_{\rm m}$ is less than $L_{\rm bol}$.
        The closest point to the dashed line is outflow No. 21, the source L1448 U-star.
        \citet{Bachiller_C90} pointed out that it was near the upper edge of the populated zone
        in the mechanical power versus stellar luminosity diagram by \citet{Lada85}.
        Now that the size of the sample is much larger,
         the closest point is still located at the upper edge.
        The lowest point represents outflow No. 337, the source GN21.38.9.
        This is the lowest-mass Bok globule according to ~\citet{Duvert_CB90}.

        In Fig.~\ref{Lbollm}, the solid line is a linear least-square fit:
            $\log L_{\rm m}= (-1.98 \pm 0.14)+(0.62 \pm 0.04)\log L_{\rm bol}$.
        The correlation coefficient is 0.69.
        One can see that the two lines are not parallel.
        The average deviation between the two lines is larger for the  high mass group
         than  the low mass ones.
        The average values of the ration $L_{\rm m} / L_{\rm bol}$  are $0.033 \pm 0.086$ and $0.0038\pm 0.017$
            for the low mass and high mass group sources, respectively.

        Figure~\ref{LbolF} is the plot of the force required to drive the outflow against the
            bolometric luminosity of the central source.
        The dashed line shows $F=L_{\rm bol}/c$.
        All the outflows are above the line,
            which means that the radiation pressure of the central sources
             would not be   enough to drive the outflows
            if the radiation photons from the central source were scattered once.
        The two plotted parameters are still correlated.
        The solid line presents the least square fit for
        the force of the flow as a function of the bolometric luminosity:
            $\log F=(-0.92 \pm 0.15)+(0.648 \pm 0.043)\log L_{\rm bol}$.
        The correlation coefficient is 0.72.

        \citet{Bally_L83} mapped a sample of sources using a
        single telescope; they
            discovered a weak correlation between the mechanical luminosity,
            the required driving force of the outflow and the
            bolometric luminosity of the central source.
        The correlation means that the energy and momentum of an outflow is determined
            by the central luminosity or mass and the physical driving engines are similar
            for all the sources \citep{Bally_L83}.
        The above authors also pointed out that the scatter of the plots
        could be understood as the result of several selection effects.
        The uncertainties in determining luminosities of central sources can cause significant scattering.
        A large part of the scatter may be due to the geometrical velocity determination.

        To test the correlation between the mechanical luminosity, the force and the luminosity of the central source,
            the increased sample size is expected to ensure that the scatter in $L_{\rm m}$-$L_{\rm bol}$ plane
            will be more than that in the $F$-$L_{\rm bol}$ plane \citep{Bally_L83}.
        Here we have a sample 25 times the size  of theirs;
            the correlation coefficient of the line fit for the $F$-$L_{\rm bol}$ (0.72)
            is just slightly  better than that for $L_{\rm m}$-$L_{\rm bol}$ (0.69).
        The standard deviation from the fitted line of $F$-$L_{\rm bol}$
            is about the same as that of $L_{\rm m}$-$L_{\rm bol}$
            or without a significant difference.
        This may be caused by various uncertainties in our sample,
        which are mentioned in Sect. 4, while
        the sample of   \citet{Bally_L83}
        came from a single survey.
        The parameter ranges of our sample are wider than
        the sample in \citet{Bally_L83}.
        For example, outflow mass is $10^{-3}\sim 10^3 \,{\rm M_{\odot}}$,
        while theirs is from $0.1\sim 100\,{\rm M_{\odot}}$.

        The mass entrainment rate ($M/t$) is also
        correlated linearly to  $L_{\rm bol}$ ~\citep{Wu_04}.
        This is further evidence showing the dependence of outflow on its driving source.

    \subsection{Dynamic time\label{tds}}

        \subsubsection{The average age}

            The dynamic time scale, $t$, is available for 275 sources.
            The time scales range from 400 yr \citep[see NGC2024 FIR6,][]{Richer90}
            to $6\times 10^5 \, \rm yr$ \citep[see L1551NE,][]{Moriarty-Schieven_W91}.
            The average value is $9.8\times 10^4 \, {\rm yr}$
            and $5.0\times 10^4 \, {\rm yr}$
            for the high mass and low mass group respectively.
            To investigate why the dynamical timescale of high
            mass sources appears longer than that of low mass sources,
            we examined the sources with $t>2\times 10^5$~yr.
            There are 21 such sources, among which 14 belong to the high mass group
            and 7 belong to the low mass group.
            We list all the sources with $t > 2\times 10^5~\rm yr$ in Table 2.

            Among 14 high mass sources, 10 are UC HII or bright far infrared sources,
            which are more evolved among massive YSOs.
            Another reason why the high mass group has a  higher average age
            is the observation resolution:
            14 of the above 21 sources have large distances $D>1.0$ kpc,
             while medium (10$\sim$15m)
            or small telescopes (3$\sim$5m) were employed
            to observe 13 of them;
            among these 13 sources, 12 are high mass sources.

        \subsubsection{The relation between outflow mass and dynamic time}

\begin{figure}
    \centering
    \resizebox{\hsize}{!}{\includegraphics[]{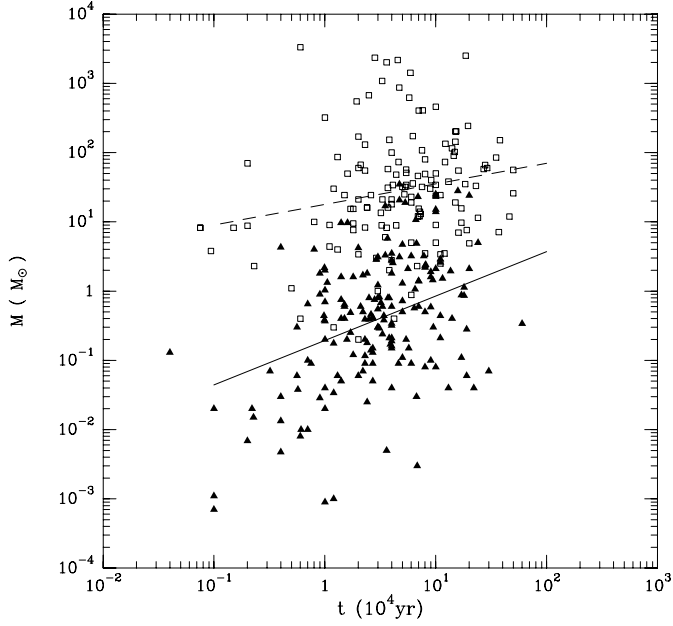}}

        \caption{Outflow masses $M$ versus dynamical time.
            The symbols are the same as for Fig.~\ref{Rcoll}.
            The dashed and solid lines are the least square linear fit for sources
            of the high and low mass group respectively.}
     \label{Mt}
\end{figure}

            The plot of the outflow mass against the dynamic time is shown in
            Fig.~\ref{Mt}.

            For 98.5\% (271/275) of the sources the dynamic timescales
                are between $10^3\sim 5.5\times 10^5$ years,
            including 99\% (155/157) of the low mass sources,
                and 98\%(116/118) of the high mass ones. 
            For the low mass group, the flow mass seems to increase with time
                while there is no such simple trend for the high mass group.
            Linear fitting shows $\log M = (1.25\pm0.11)+(0.30\pm 0.12)\log t$,
                and a poor $r=0.20$. The correlation is much
                stronger for the low mass group, $\log M = (-0.71\pm 0.08)+(0.64\pm 0.11)\log t$,
                and $r=0.41$. Thus the flow processes may be different for the two groups.

\section{Associated phenomena \label{associated}}

     Numerous associated objects were detected and investigated.
        Ninety (23\%) sources have associated HII or centimeter continuum emission regions,
            136 (35\%) have optical jets or HH objects,
            92 (24\%) have H$_2$ jet and 190 (49\%) have H$_2$O masers.
        Among 136 sources with HH objects or optical characters, 14 (10\%) are in the high mass group.
        In the 190 sources with water masers, 75 (39\%) are in the low mass group.
        There are 45 (12\%) outflows with collapses
        (the percentage in the high mass group is 20\%),
        nearly all of the collapse candidates detected so far.
        According to~\citet{Shu_AL87}, the collapse should occur earlier than bipolar outflow 
        in star formation stages. However, nearly all of the
        collapses were detected in outflow sources. This may be because
        the infall material near the embedded sources is shielded by surrounding gas and dust, thus is difficult
        to observe until the outflow later expels the surrounding
        material so that the internal region is observable.  


\section{Spatial distribution and occurrence rate\label{distri}}

    The three dimensional spatial distribution of the sources
    projected on the coordinate plane is given in Fig.~\ref{xyz}.
    The sun is at the origin, with the X axis pointing to the 90 degree longitude direction of the IAU Galactic coordinate,
       the Y axis passing through the Galactic Center and the Z axis pointing to the north Galactic pole.
        High mass sources are represented by clear squares and low
        mass ones by filled triangles.
    Figure~\ref{xyz}a shows the sources projected on the Galactic plane;
    Fig.~\ref{xyz}b is the same frame as Fig.~\ref{xyz}a
    but shows only the inner 2 kilo-parsecs from the sun.
    Figure~\ref{xyz}c and Fig.~\ref{xyz}d are projections onto the X-Z and Y-Z planes, respectively.

    A prominent feature seen in the plots
    is that there are more sources between $l = 30^{\circ}$
    to $l = 210^{\circ}$
        than in the rest ($\sim 2:1$; see Table 1a).
    This is largely attributed to the fact that most observations were
        performed on sources that can be seen in the northern hemisphere.
    There do exist spatial structures that are asymmetrically distributed with respect to the sun
        in the local distance, e.g. the neutral interstellar gas
        \citep{Frisch_Y83}.
    Such asymmetry mirrors the distribution of the nearby O and B stars associated with the Gould Belt \citep{Stothers_F74}.
    The contour of $N({\rm H})\sim 5\times 10^{20} {\rm cm^{-2}}$ is shown in Fig.~\ref{xyz}a
    as a thick line \citep{Frisch_Y83}.
    A molecular disk model with radius 1.5kpc and a molecular ring with radius 3-7 kpc
        are also given in Fig.~\ref{xyz}a \citep{Liszt_B78, Scoville_S87}.
    Multiple components along the line of the sight towards these molecular rings where $|\, l \, |$ is less than $~40^{\circ}$
        may be another reason that makes it difficult to detect or to identify  molecular outflows in these regions.

    There are 230 sources in the catalogue within one kilo parsec of the sun.
    The occurrence rate can be estimated as $74/t \,{\rm kpc^{-2}yr^{-1}}$,
        where t is the average lifetime (in years) of the outflows.
    For $t = 5.8\times 10^4$ years, which is the average value of the 230 sources within one kiloparsec,
        the birth rate is $1.3\times 10^{-3}\,{\rm
        kpc^{-2}yr^{-1}}$.

    Among the 230 sources within 1 kpc of the sun there are 171 sources with available bolometric luminosity (see Table~\ref{birth}).
    Taking the upper and lower limit of masses of the evolution tracks on the observed H-R diagram according to
        bolometric luminosity \citep[][ and references therein]{Levreault85},
        the average mass of these sources was estimated to be $3.6{\rm M_{\odot}}$ as an upper limit,
        and $2.1{\rm M_{\odot}}$ as a lower limit (see Table~\ref{birth}).

    The birth rate measured for the number of stars,
    $1.3\times 10^{-3}~{\rm kpc^{-2}yr^{-1}}$,
    is equal to $4.7 \times 10^{-9}~{\rm M_{\odot}pc^{-2}yr^{-1}}$
       measured as a stellar mass upper limit;
        for the lower limit, it is $2.7 \times 10^{-9}\,{\rm M_{\odot}pc^{-2}yr^{-1}}$ (see Table~\ref{birth}).
    Both are similar to the local star formation rate $3\sim 6 \times 10^{-9}\,{\rm M_{\odot}pc^{-2}yr^{-1}}$,
        or $4\sim 11 \times 10^{-9}~{\rm M_{\odot}pc^{-2}yr^{-1}}$,
        depending on whether there is any compression in the spiral shock wave or not~\citep{Kaufman79}.
    These results suggest that high-velocity outflows are common in star formation.

\section{Summary\label{sum}}

    In this paper, the searches of molecular outflow sources are reviewed.
    Our Catalogue includes 391 high velocity molecular outflows and their basic parameters.
    We  classified them into two groups according to the
    available bolometric luminosity or to the outflow mass.
    We analyzed the progress of molecular outflow searches and
    divided the history of outflow search into four periods:
    I.1976-1984, II. 1985-1989,
    III. 1990-1995 and IV. 1996-2003.
    The characters of different periods and the developmental
    trend are discussed.
    The morphology, collimation and physical parameters of outflows  were
    investigated.
    We also studied the correlation between outflows and their
    central sources.
    Finally the spatial distribution and birth rate of young stars
     are presented and discussed.
    For convenience, the main statistical results are listed in Table. 4.

\begin{figure}
    \centering
    \resizebox{\hsize}{!}{\includegraphics[]{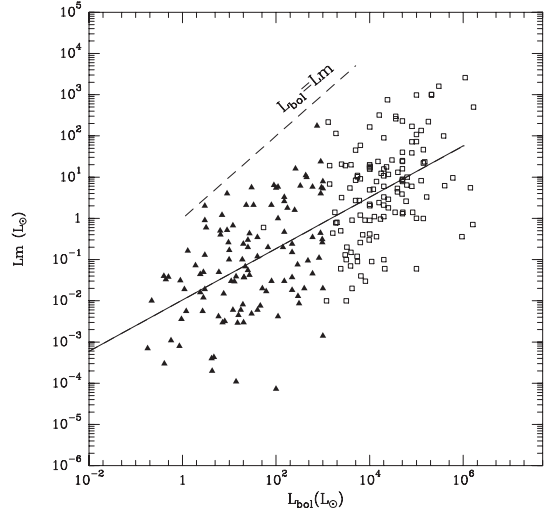}}

        \caption{The outflow luminosity ($L_{\rm m}$) versus bolometric luminosity ($L_{\rm bol}$).
            The symbols are the same as for Fig.~\ref{Rcoll}.
            The relation $L_{\rm m} = L_{\rm bol}$ is shown as a dashed line.
            The solid line is the least square linear fit line.}
     \label{Lbollm}
\end{figure}

\begin{figure}
    \centering
    \resizebox{\hsize}{!}{\includegraphics[]{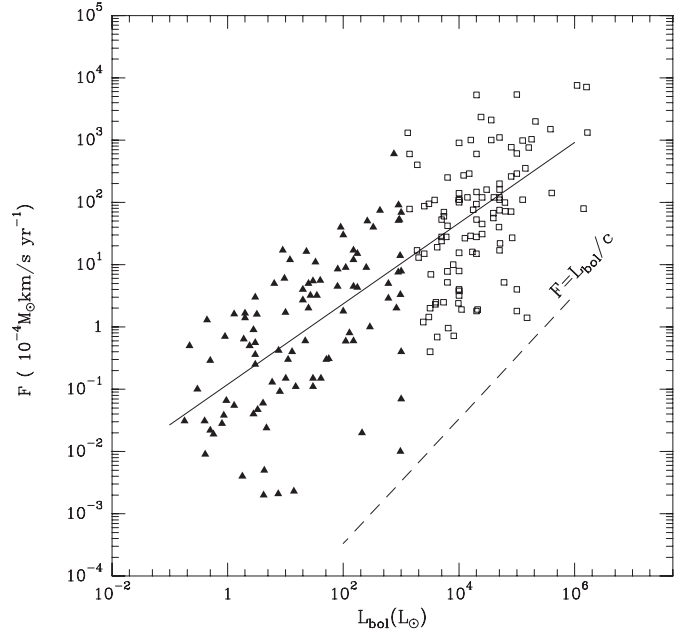}}

        \caption{Outflow force $F$ versus bolometric luminorsity ($L_{\rm bol}$)
            of the associated infrared souces.
            The symbols are the same as in Fig.~\ref{Rcoll}.
            The dashed line presents the relation $F=L_{\rm bol}/c$.
            The solid line is the least square linear fit.}
     \label{LbolF}
\end{figure}

    Our main results are as follows:

    \begin{enumerate}
    \item The catalogued outflows have been classified into high mass and
          low mass groups
          according to the bolometric luminosity of the center sources
          and the outflow masses.
          The pace of outflow identification has accelerated
          in the past two decades.
          The number of outflows detected  per
          year tends to increase, that is, 7 in Period I but 23 in
          Period IV.
          The number of low mass sources increased fast during period II.
          In period IV, the number of high mass outflows increased rapidly,
          suggesting that the outflow phenomenon is also common
          in high mass star formation regions.
          Data acquired recently suggest  high mass star formation processes
          may also go through mass accreting and ejecting,
          similar to those processes in low mass star formation.

    \item Polarity was known for all the 391 sources, 84\% of which are bipolar.
          Red mono-polar outflows are found to be  equally abundant
          compared to the blue ones within the statistical uncertainty.
          This suggests that  monopolarity of the outflows
          is intrinsic and is a result of the driving source
          being partially submerged in the molecular cloud.
          The number of multiple sources has increased
          and isolated ones decreased
          since the last catalogue \citep{Wu_HH96},
          which shows the improvement of the detection technology.

    \item We analyzed the collimation of the outflows.
          The number of sources with known collimation factors has increased
          from 26 in  period I to 213 in  period IV.
          The current average value of the collimation factors
          is slightly larger than that in period I.
          Although the resolution of the equipment
          used in period IV has generally improved,
          the number of high mass outflows observed increased in this period.
          As high mass outflows have in general lower collimation
          factor values, the average collimation value only changed slightly.

    \item The collimation of low mass sources is better than that of
          high mass ones as a whole.
          Different outflow driving mechanisms may explain such differences.
          Moreover, projection effects, physical conditions of the environment and
          evolutionary states also affect the morphology of outflows.

    \item The relationship between  physical parameters of the outflow
          and its central source is examined.
          Analysis shows that outflow masses are correlated strongly
          with bolometric luminosities of young stellar sources.
          This relation, a typical example of an indirect
          correlation of two quantities, is obtained for the first time.

    \item Similar linear correlations exist
          between flow mechanical luminosity,
          the force required to drive the outflow
          and the bolometric luminosity in logarithmic scales.
          The weak correlation found by \citet{Bally_L83} in an early survey is confirmed.

          The mass, momentum and energy of the outflows depend on
          the luminosity of the central source.
          These relationships between the parameters of the outflows
          and the driving sources suggest that the
          energetic outflows originate from their central source.
          Although the radiation pressure could not drive the outflow directly
          if the photons emitted from the central source  scatter once,
          these correlations confirm the  relationship
          between outflows and their central stellar objects.

    \item Outflow mass and dynamic time tends to be correlated for low mass
          sources while for high mass ones there was no such correlation.
          This together with the differences between
          the low mass and high mass group  sources shown by  $L_{\rm m}$ v.s. $L_{\rm bol}$ (Fig.
          \ref{Lbollm}) and
          $F$ vs. $L_{\rm bol}$ (Fig. \ref{LbolF})
           is consistent with the presence of  different driving mechanisms.

    \item Objects associated with outflows were investigated.
          136 outflows have associated  HH objects; 10\% of those
          are among the high mass group; 190 outflows have water masers,
          39\%  belong to the low mass group.
          12\% of the
          outflow sources  have ongoing collapse of infall detected.

    \item The spatial distribution of known outflows is not symmetric
          with respect to the sun
          because the southern hemisphere has fewer identified sources,
          largely due to less observation.
          The occurrence rate within 1kpc from the sun is
          $74/t ~{\rm kpc^{-2}yr^{-1}}$,
          where $t$ is a typical lifetime of outflow in years.
          For an outflow dynamic time of $5.8\times 10^4$ years
          the birth rate is $1.3\times 10^{-3}~{\rm kpc^{-2}yr^{-1}}$,
          close to the local star birth rate.
    \end{enumerate}

    Higher angular resolution and sensitivity are necessary
        to further resolve the velocity structure and survey the distant sources.
    If a large number of outflows with more accurate morphology are found,
        we may better understand the physical nature of the energetic processe
        occuring during
         early stellar evolution.


\begin{acknowledgements}

     We would like to thank the anonymous referee for
     constructive suggestions and helpful comments.

     We are very grateful to J. Wu, J. Wang, C. Yang,  M. Wei, W. Zhou and L. Zhu
     for their assistance in the data collection and data checking.
     We thank Hongping Du and the editor for help with our English expressions.
     This project is supported by the grants
     from NSFC No. 10133020, 10128306 and 10203003,
     and by G1999075405 of NKBRSF.

\end{acknowledgements}

\bibliography{wu} 

\newpage

\setcounter{table}{1}

\begin{table}
    \begin{flushleft}
    \caption{Outflow with dynamic time $>20\times 10^4 \rm yr$}
    \label{longtime}
    \begin{tabular}{ccccccc}
    \hline
    {\parbox[t]{20mm}{\centering {Outflow\\No.}}}&%
    {\parbox[t]{30mm}{\centering {Outflow\\Name}}}&%
    {\parbox[t]{12mm}{\centering {$D$\\kpc}}}&%
    {\parbox[t]{15mm}{\centering {$t$\\ $10^4~\rm yr$}}}&%
    {\parbox[t]{20mm}{\centering {Group}}}&%
    {\parbox[t]{20mm}{\centering {Telescope\\used}}}&%
    {\parbox[t]{14mm}{\centering {Ref.}}}\\
    \hline
    009 & 00494+5617 & 2.2  & 24  & H & 14m          & 3    \\
    014 & 02310+6133 & 3.8  & 51.5& H & 13.7m        & 381  \\
    015 & 02461+6147 & 4.2  & 35  & H & 3m($J=3-2$)  & 380  \\
    040 & B5-IRS4    & 0.35 & 24  & L & 7m           & 21   \\
    052 & L1551NE    & 0.16 & 60  & L & 12m          & 128  \\
    126 & 05490+2658 & 2.1  & 37  & H & 14m          & 3    \\
    130 & 05553+1631 & 2.5  & 46  & H & 14m          & 3    \\
        & G192.16    & 2.0  & 27  & H & 12m          & 194  \\
    131 & CB39       & 0.6  & 30  & L & 14m          & 178  \\
    139 & S255       & 2.5  & 50  & H & 14m          & 117  \\
    162 & 07028-1100 & 1.4  & 24  & L & 13.7m        & 421  \\
    166 & WB89 1135  & 6.42 & 37.9& H & 15m          & 196  \\
    173 & WB89 1275  & 6.38 & 27.8& H & 15m          & 196  \\
    237 & 18150-2016 & 1.8  & 24  & H & 4m           & 260  \\
    281 & 19471+2641 & 2.62 & 21  & H & 10m($J=3-2$) & 203  \\
    325 & 21015+6757 & 0.44 & 24  & L & 3m($J=3-2$)  & 397     \\
    336 & V645 Cyg   & 5.6  & 50  & H & 5m($J=2-1$)  & 8    \\
    337 & GN21.38.9  & 0.75 & 30  & L & 10m          & 150  \\
    350 & 21432+4719 & 1.0  & 22  & L & 45m          & 258  \\
    360 & 22142+5206 & 4.5  & 23  & H & 45m          & 187  \\
    374 & 22475+5939 & 4.7  & 29  & H & 12m          & 199  \\
    382 & MBM55      & 0.18 & 22  & H & 12m          & 100  \\
    \hline
    \end{tabular}
    \end{flushleft}
\end{table}

\begin{table}
    \begin{flushleft}
    \caption{$L_{\rm bol}$ and mass ranges for associated stellar sources of outflow within 1 kpc.}
    \label{birth}
    \begin{tabular}{c|ccccccc}
    \hline
    $L_{\rm bol}({\rm L_{\odot}})$&%
    $<10^{-1}$&%
    $10^{-1}\sim 10^{0}$&%
    $10^{0}\sim 10^{1}$&%
    $10^{1}\sim 10^{2}$&%
    $10^{2}\sim 10^{3}$&%
    $10^{3}\sim 10^{4}$&%
    $>10^{4}$\\
    \hline
    Source Number & 0 & 20 & 50 & 49 & 36 & 8 & 8 \\
    Upper limit $\rm M_{\odot}$ & 1.0 & 1 & 1.5 & 3 & 5 & 9 & 15 \\
    lower limit $\rm M_{\odot}$ & 0.2 & 0.5 & 1.0 & 1.5 & 3 & 5 & 9 \\
    \hline
    \end{tabular}
    \end{flushleft}
\end{table}

\clearpage

\begin{table}
    \begin{flushleft}
    \caption{Index of main statistical results}
    \label{longtime}
    \begin{tabular}{lccc}
    \hline
    {\parbox[t]{40mm}{\centering {Item}}}&%
    {\parbox[t]{40mm}{\centering {Result}}}&%
    {\parbox[t]{15mm}{\centering {Section}}}&%
    {\parbox[t]{14mm}{\centering {Paragraph}}}\\
    \hline
    Outflows cataloged& 391+6 & 2.1  & 2    \\
    High group ($L_{\rm bol}>10^3{\L_\odot}$ or $M>3{\rm M_\odot}$)& 139 & 2.3& 2  \\
    low group ($L_{\rm bol}\leq 10^3{\L_\odot}$ or $M\leq 3{\rm M_\odot}$)& 223 & 2.3  & 2  \\
    Sources detected in Period I, II, III, IV& 55, 95, 88, 159     & 2.4 & 2  \\
    High group in Period I, IV& 31\%, 38\%      & 2.4 & 3  \\
    Sources with $L_{\rm bol}>10^4$ to ${\rm L_{\odot}}$ & 23\% & 2.4  & 4  \\
    Sources with UC HII regions & 23\% & 2.4  & 4  \\
    Bipolar sources  & 327(84\%)& 3  & 1  \\
    Monopolar sources & 50(13\%) [28 red, 22 blue]& 3  & 1  \\
    Multipolar sources  & 12(3\%)& 3  & 1  \\
    Isotropic sources  & 2& 3  & 1  \\
    Sources with available R$_{\rm coll}$& 213& 3  & 3  \\
    Average $R$$_{\rm coll}$& 2.45$\pm$ 1.74& 3  & 3  \\
    Average $R$$_{\rm coll}$ of high group& 2.05$\pm$ 0.96& 3  & 3  \\
    Average $R$$_{\rm coll}$ of low group& 2.81$\pm$ 2.16& 3  & 3  \\
    Bolometric luminosity range& 0.1-$10^6$$\rm L_{\odot}$& 2.3  & 1    \\
    Uncertain factor of $M$& 6.4 & 4  & 2  \\
    Uncertain factor of $F$& 6.7 & 4  & 2  \\
    Uncertain factor of $L$$\rm _m$ & 7.1 & 4  & 2  \\
    $R_{\rm max}$ range& 0.01 to 3.98 pc& 4  & 1  \\
    Mass range& $10^{-3}$ to $10^3 \,\rm M_{\odot}$& 4  & 1  \\
    Momentum & $10^{-3}$ to $10^4 \rm \,M_{\odot} km/s$& 4  & 1  \\
    Energy & $10^{38}$  to $10^{48}$ ergs& 4  & 1  \\
    Mechanical luminosity&$10^{-5}$ to $10^3 \,\rm L_{\odot}$ & 4  & 1  \\
    Force $F$ & $10^{-7}\sim 10^{2}$ $\rm M_\odot km/s\,yr$ & 4 & 1 \\
    Mass loss rate $\dot{M}$ & $10^{-9}\sim 10^{-3}$ $\rm M_\odot /yr$& 4 & 1 \\
    Dynamical time available sources & 275 & 4.3.1  & 1  \\
    Range of dynamical time & $4\times 10^2$ to $6\times 10^5$ yr & 4.3.1 & 1\\
    Dynamical time of  the 99\% sources & $10^3\sim 5.5\times 10^5$ yr & 4.3.2  & 2 \\
    Average dynamical time of high group& $9.8\times 10^4 \, {\rm yr}$ & 4.3.1  & 1  \\
    Average dynamical time of low group& $5.0\times 10^4 \, {\rm yr}$ & 4.3.1  & 1  \\
    Sources with $t>2\times 10^5$~yr& 21 [14 in high  group] & 4.3.1  & 1  \\
    Sources associated with HH objects& 136 (35\%) [90\% in low group]& 5  &  \\
    Sources associated with H$_2$ jet & 92 (24\%)& 5  &   \\
    Sources associated with H$_2$O masers& 190 (49\%) [115 (61\%) in high group]& 5  &   \\
    Collapse candidates & 45 (12\%)& 5  &   \\
    Sources within 1 kpc& 230  & 6  &3   \\
    Sources within 1 kpc ($L_{\rm bol}$ available) & 171 [16(9\%) in high group] & 6 & 4\\

    \hline
    \end{tabular}
    \end{flushleft}
\end{table}

\clearpage

\begin{figure}
    \centering
    \resizebox{\hsize}{!}{\includegraphics[]{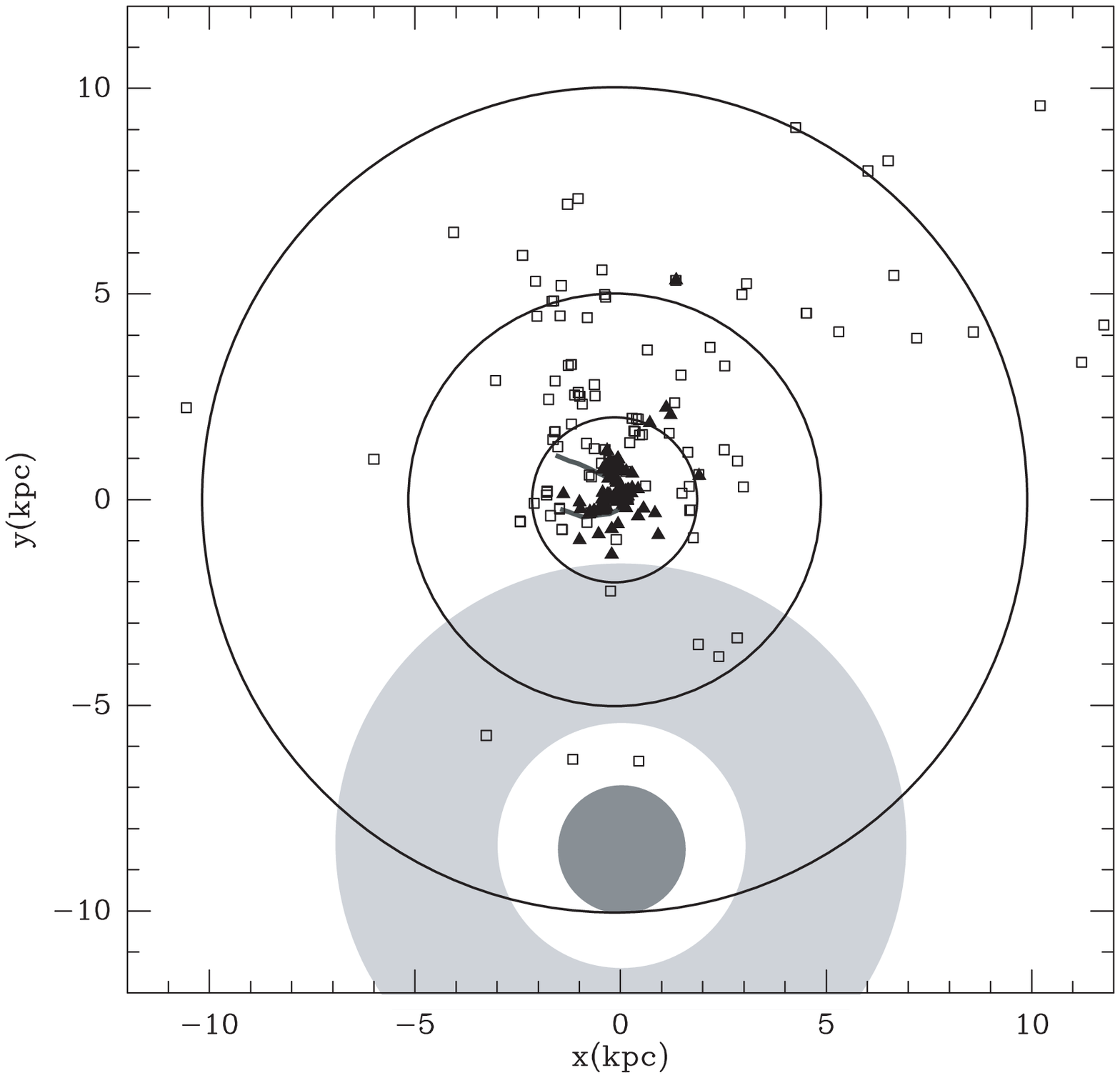}}
    { \large a}
    \resizebox{\hsize}{!}{\includegraphics[]{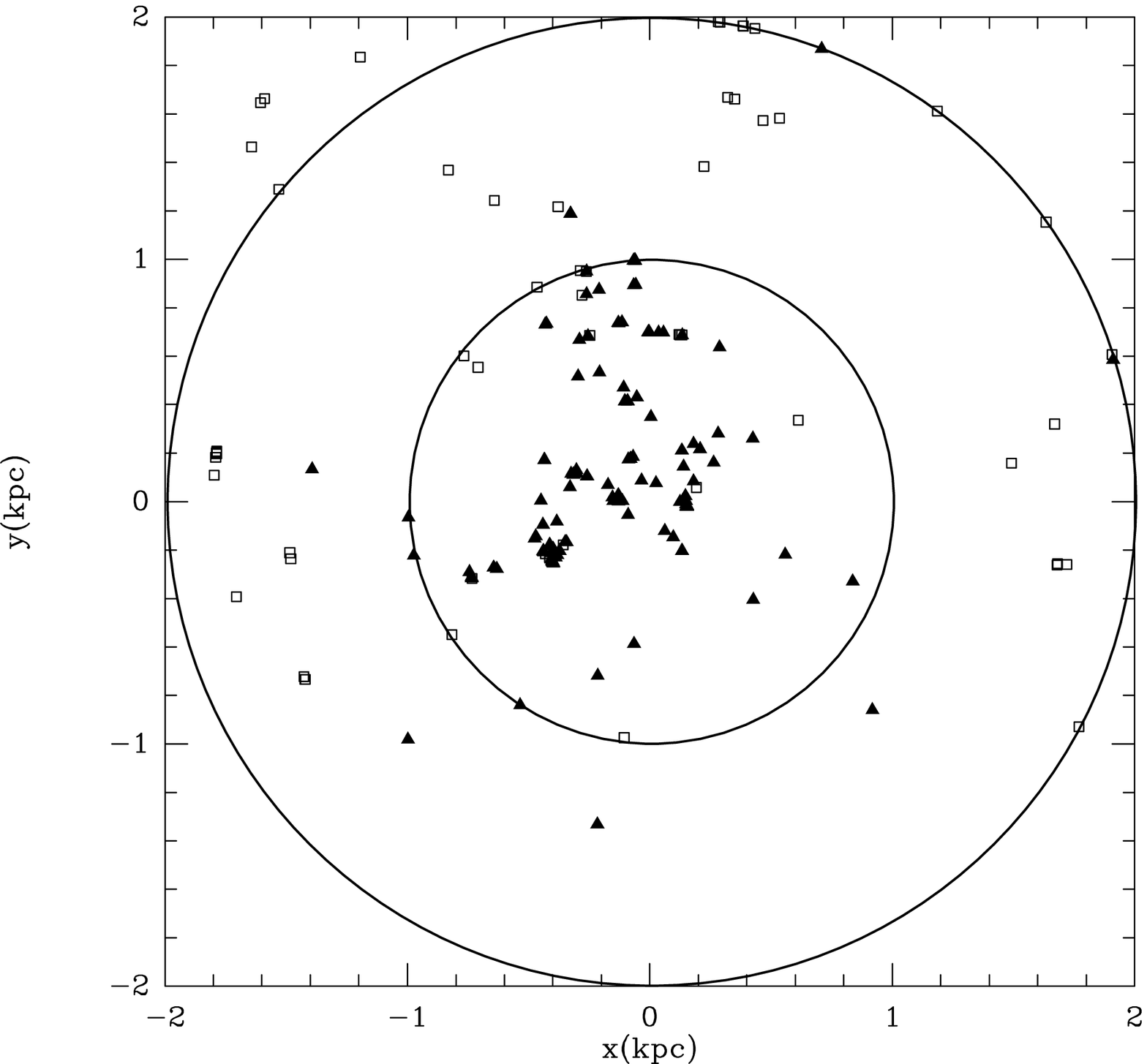}}
    {\large b}

    \caption{Spatial distribution of outflows.
        The sun is at the origin of the coordinate.
        The X-axis points to the $90^{\circ}$ longitude direction
            of the IAU Galactic coordinate, the Y-axis passes through Galactic Center and
            the Z-axis points to the north Galactic pole.
            Triangles represent  low mass outflows and circles  high mass ones.
            The thick dark grey line with in the 2 kpc radius circle denotes the contour of
            $N({\rm H})\sim 5\times 10^{20}~{\rm cm^{-2}}$
            ~\citep{Frisch_Y83}.
        a). The outflow distribution on the coordinate plane x-y.
        The three circles from inner to outer indicate
            distances
            of 2 kpc, 5 kpc and 10 kpc from the sun.
            The small shadow elliptic presents the projection of a molecular disk model
            with a radius of 1.5 kpc~\citep{Liszt_B78}.
            The shadow ring indicates projection of the molecular ring with radius
            3$\sim$7 kpc~\citep{Scoville_S87}.
        b). The  same as a), but shows only the inner 2 kpc from the sun.
        The two circles indicate 1 kpc and 2 kpc  from
        the sun.
        c). Projection onto the X-Z plane.
        d). Projection onto the Y-Z plane.
        Note that the vertical axes in c) and d) are scaled up for clarity. }
         \label{xyz}
\end{figure}

 \addtocounter{figure}{-1}

\begin{figure}
    \centering
    \resizebox{\hsize}{!}{\includegraphics[]{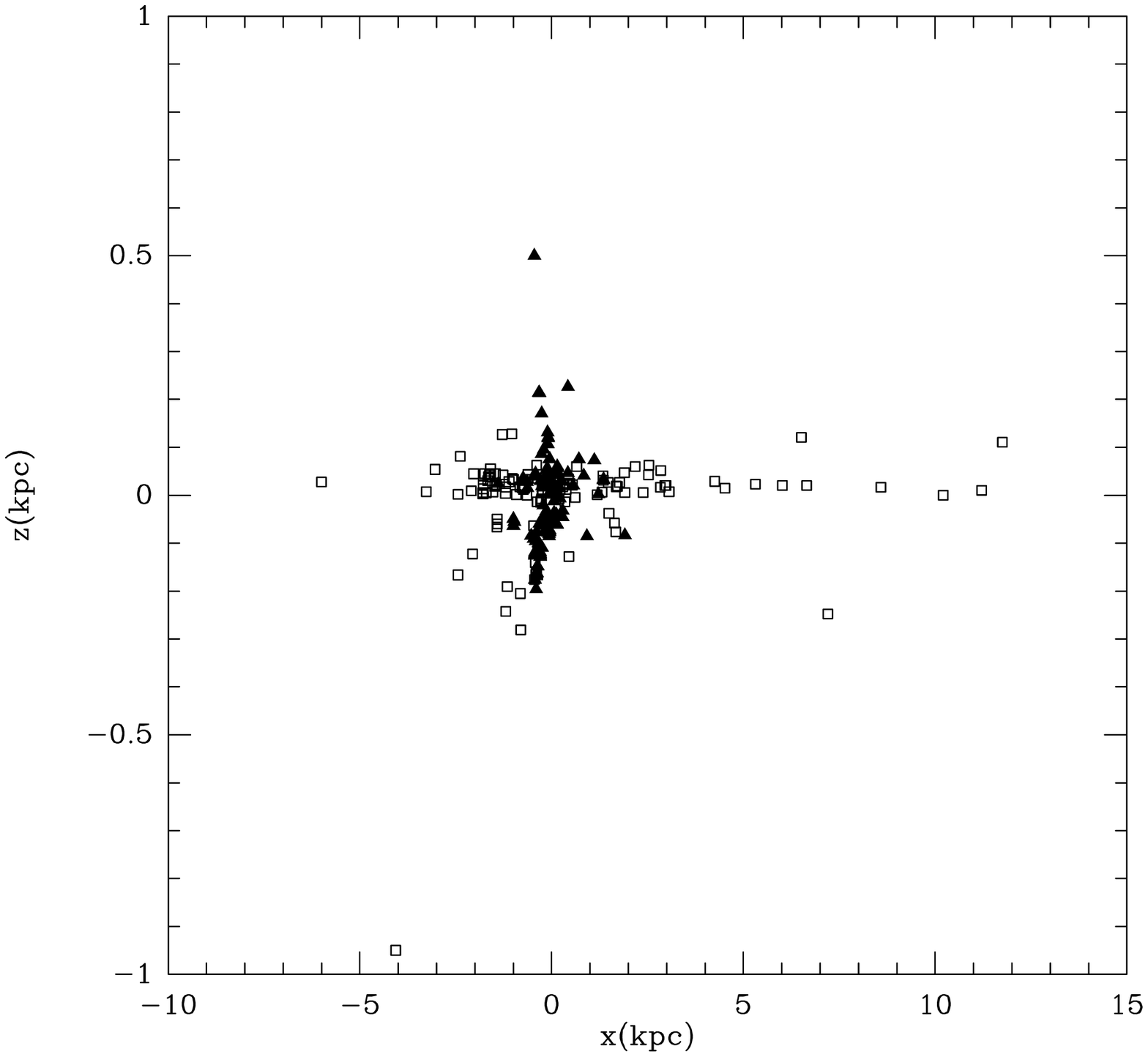}}
    {\large c}
    \resizebox{\hsize}{!}{\includegraphics[]{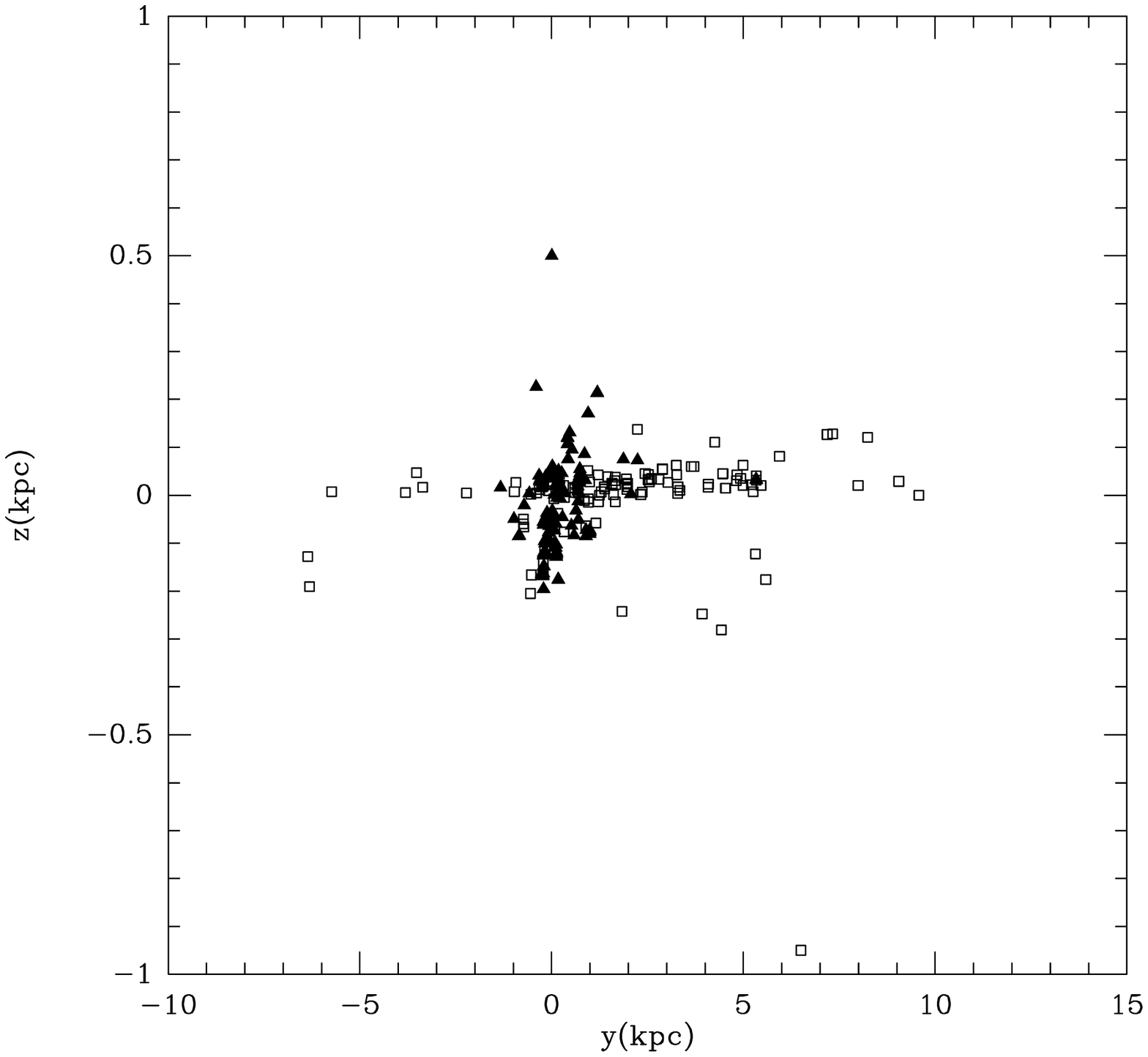}}
    {\large d}
    \caption{\textit{continued}}
    \label{xyz3}
\end{figure}

\end{document}


\setcounter{page}{21} \columnseprule 0.1mm \columnsep 20mm
\noindent a `II' indecates HII region; `H' indicates optical jet
or HH object; `2' indicates H$_2$ jet; `W' indicates water maser;
`I'
indicates infall or collapse.\\
b The maps of these sources are not complete or confused with other sources.\\
c These sources are detected with CO J=3-2 or higher.\\
d These outflows without corresponding young stellar objects.\\
e The underlined R is quoted from references. \\
f This source includes NGC7538 IRS1, IRS9, IRS11, and A.\\
g The distance is one of the assumed values(cf. Ref. 123).\\
h The mass is calculated under optical thick.\\
i This reference proposes that 05358+3543 consists of CO flow A, B
and a SiO flow C. \\
j The authors of Ref. 326 proposed that this source has 2
overlapping outflows.\\
k The HH objects detected before 1999 were all from Ref. 251.\\
l This source includes RNO43S and RNO43N. \\
m The parameters are derived from C$^{18}$O observation. \\
n This source includes IC1396N BIMA1 and BIMA2.\\
o $\dot M_f$ is not used if it is deduced by $M/\tau$.\\
p It was not identified as bipolar outflow by Ref. 191.\\
q Ref. 327 reported that there may be two outflows.\\
r Its reshift lobe is probably masked by the outflow of L379
IRAS(3).\\
s This timescale is calculated with the assumption that the clump
has a tangential velocity of 200 km/s.\\
t This source includes NGC2024 Fir 4, 5 and 6.\\
u This source includes NGC7129 FIRS2-MM1, FIRS2-IR, FIRS1-IRS6 and
FIRS1-MM1.\\
v Parameters calculated here are not for the whole outflow but only for the n4 region.\\
w The author of Ref. 354 proposed that outflows at this position
are not driven by FIR sources but by jets from faint undetected
yet low-mass stars; the parameters in Table 1a are the total of
three outflows identified.\\
 \twocolumn
\begin{bf}
{\large \bf References\\}
\end{bf}
1, Bally \& Lada, 1983; \\
2, Cant\'o et al., 1984; \\
3, Snell et al., 1990; \\
4, Brackmann \& Scoville, 1980;\\
5, Arquilla \& Goldsmith, 1984; \\
6, Nakano \& Yoshida, 1986; \\
7, Lada, 1985; \\
8, Levreault, 1985;\\
9, Bachiller et al., 1990b; \\
10, Lada \& Harvey, 1981; \\
11, Snell et al., 1984; \\
12, Levreault, 1988; \\
13, Mao et al., 1989; \\
14, Goldsmith et al., 1984;\\
15, Yamashita et al., 1989a; \\
16, Phillips et al., 1988;\\
17, Edwards \& Snell, 1984; \\
18, Edwards \& Snell, 1983; \\
19, Snell \& Edwards, 1981; \\
20, Beichman et al., 1984; \\
21, Goldsmith et al., 1986; \\
22, Myers et al., 1988; \\
23, Myers et al., 1987; \\
24, Heyer et al., 1987; \\
25, Snell et al., 1980; \\
26, Strom et al., 1976; \\
27, Lichten, 1982a; \\
28, Beichman et al., 1986;\\
29, Liljestr\"om et al., 1989;\\
30, Frerking \& Langer, 1982; \\
31, Wu et al., 1992; \\
32, Snell et al., 1988;\\
33, Fukui et al., 1986; \\
34, Morgan \& Bally 1991; \\
35, Ziurys et al., 1990; \\
36, Wouterloot \& Brand, 1989;\\
37, Schmid-Burgk et al. 1989; \\
38, Erickson et al., 1982; \\
39, Kwan \& Scoville 1976; \\
40, Olofsson et al., 1982; \\
41, Zuckerman et al., 1976; \\
42, Hilton et al., 1986; \\
43, Snell \& Edwards, 1982;\\
44, Campbell \& Thompson, 1984;\\
45, Bally, 1982; \\
46, Lichten, 1982b;\\
47, Margulis et al., 1988; \\
48, Haschick et al., 1981; \\
49, Rodriguez et al., 1982; \\
50, Blitz \& Thaddeus, 1980; \\
51, Lada \& Gautier, 1982; \\
52, Levreault, 1983; \\
53, Margulis \& Lada 1986;\\
54, Dent et al., 1984; \\
55, Rodriguez et al., 1980;\\
56, Tamura et al., 1990; \\
57, Fischer et al., 1982; \\
58, Simon et al., 1984; \\
59, Mozurkewich et al., 1986; \\
60, Ho et al., 1982; \\
61, Armstrong \& Winnewisser, 1989;\\
62, Edwards \& Snell, 1982;\\
63, Little et al., 1988; \\
64, Lada et al., 1984; \\
65, Dickel et al., 1978; \\
66, Levreault 1984; \\
67, White \& Phillips 1981; \\
68, Sato \& Fukui 1989; \\
69, Mathieu et al., 1988; \\
70, Rodr\'\i guez et al., 1981;\\
71, Haikala \& Laureijs, 1989;\\
72, Dent et al., 1985a; \\
73, Dent et al., 1985b; \\
74, Cant\'o et al., 1981; \\
75, Reipurth \& Olberg, 1991;\\
76, Kameya et al., 1989; \\
77, Evans et al., 1981a; \\
78, Harvey et al., 1984; \\
79, Schwartz et al., 1983; \\
80, Calvet et al., 1983; \\
81, Wolf et al., 1990; \\
82, Hayashi et al., 1987;\\
83, Evans et al., 1981b; \\
84, Terebey et al., 1989; \\
85, Yang et al., 1990; \\
86, Fukui, 1989; \\
87, Arquilla \& Kwok 1987;\\
88, Fischer et al., 1985; \\
89, Sugitani et al., 1989; \\
90, Casoli et al., 1986; \\
91, Schwartz et al., 1988; \\
92, Loren, 1989; \\
93, Parker et al., 1988; \\
94, Harvey \& Forveille 1988;\\
95, Scoville et al., 1986; \\
96, Bachiller \& Cernicharo, 1990; \\
97, Xie \& Goldsmith, 1990; \\
98, Clark, 1986; \\
99, Wouterloot et al., 1989; \\
100, Vall\'ee \& Avery, 1990; \\
101, Sanders \& Willner, 1985;\\
102, Richer, 1990; \\
103, Schmid-Burgk et al., 1990; \\
104, Yang et al., 1991; \\
105, Mizuno et al., 1990; \\
106, Andre et al., 1990; \\
107, Richer et al., 1989;\\
108, Schwartz, et al., 1985; \\
109, Davidson \& Jaffe, 1984; \\
110, Iwata et al., 1988; \\
111, Olberg et al., 1992; \\
112, Mattila et al., 1989; \\
113, Smith \& Fischer, 1992; \\
114, Reipurth, 1989; \\
115, Reipurth \& Graham, 1991; \\
116, Wilking et al.,1990; \\
117, Heyer et al., 1989; \\
118, Evans et al., 1994; \\
119, Koo et al., 1994; \\
120, Phillips \& Mampaso 1991; \\
121, Bachiller et al., 1991; \\
122, Umemoto et al., 1992; \\
123, Odenwald \& Schwartz 1993; \\
124, Ladd et al., 1991; \\
125, Barsony, 1989;\\
126, Bachiller et al., 1990a; \\
127, Chernin \& Masson, 1991; \\
128, Moriarty-Schieven \& Wanner, 1991; \\
129, Wilson \& Mauersberger 1991; \\
130, Knee, 1992; \\
131, Wootten \& Loren, 1987; \\
132, Margulis et al., 1990; \\
133, Fukui et al., 1988; \\
134, Mendoza et al., 1990; \\
135, Liseau et al., 1988; \\
136, Walker et al., 1988; \\
137, Kutner et al., 1982; \\
138, Fukui et al., 1989; \\
139, Weintraub et al., 1993; \\
140, Loren et al., 1990; \\
141, Eiroa et al., 1992; \\
142, Greg\'orio Hetem et al., 1990; \\
143, Cohen \& Schwartz 1987;\\
144, Terebey et al., 1990; \\
145, Moriarty-Schieven et al., 1992; \\
146, Tamura et al., 1991; \\
147, Margulis et al., 1989; \\
148, Garden et al., 1991a; \\
149, Umemoto et al., 1991; \\
150, Duvert et al., 1990; \\
151, Koo, 1989; \\
152, Choi et al., 1993; \\
153, Margulis \& Snell, 1989; \\
154, Avery et al., 1990;\\
155, Olberg et al., 1989; \\
156, Reipurth et al., 1993; \\
157, Armstrong, 1989; \\
158, Cohen 1980; \\
159, Kenyon et al., 1993; \\
160, Torrelles et al., 1993;\\
161, Morgan et al., 1991; \\
162, Dobashi et al., 1993; \\
163, Minchin et al., 1993; \\
164, Kastner et al., 1994; \\
165, Bally et al., 1994; \\
166, Hasegawa et al., 1994;\\
167, White, 1993; \\
168, Parker et al., 1991; \\
169, Blake et al., 1995; \\
170, Minchin et al., 1995; \\
171, Sato et al., 1994; \\
172, Jiang et al., 2001; \\
173, Zeilik et al., 1978;\\
174, Masson et al., 1990; \\
175, Chernin \& Masson, 1992; \\
176, Haikala \& Dietrich, 1989; \\
177, Bourke, 1995; \\
178, Yun \& Clemens, 1992; \\
179, Wang et al., 1995; \\
180, Dobashi et al., 1994;\\
181, Wu et al, 1999a;                  \\
182, Xiang \& Turner, 1995;\\
183, Wouterloot et al., 1988;\\
184, Moritarty-Schieven et al., 1995;\\
185, Yonekura et al., 1998;\\
186, Dobashi et al., 1998a;\\
187, Dobashi et al., 1998b;\\
188, Bourke et al., 1997;\\
189, Reipurth et al., 1997;\\
190, Bontemps et al., 1996a;\\
191, Wu et al., 2004;\\
192, Lyder et al., 1998;\\
193, Shepherd \& Churchwell, 1996a;\\
194, Shepherd \& Churchwell, 1996b;\\
195, Correia et al., 1997;\\
196, Wouterloot \& Brand, 1999;\\
197, Chandler et al., 1996;\\
198, Wu et al., 1999b;\\
199, Yang \& Wu, 1998;\\
200, Hirano et al., 1997;\\
201, Codella \& Bachiller,1999;\\
202, Hunter et al., 1997;\\
203, Mao, 1999;\\
204, Ladd \& Hodapp, 1997;\\
205, Meyers-Rice \& Lada, 1991;\\
206, Fich \& Lada, 1997;\\
207, Zhang et al., 2001;\\
208, Henning et al., 2000a;\\
209, Gueth \& Guilloteau, 1999;\\
210, Chen et al., 1997;\\
211, Langer et al., 1996;\\
212, Schreyer et al., 1997;\\
213, Shepherd et al., 2000;\\
214, Gueth et al., 1996;\\
215, Lee et al., 2000;\\
216, Benson et al., 1984;\\
217, Shepherd \& Kurtz, 1998;\\
218, Sandell et al., 1999;\\
219, Henning et al., 2000b;\\
220, Yun et al., 2001;\\
221, Sugitani \& Ogura, 1994;\\
222, Knee \& Sandell, 2000;\\
223, Cabrit et al., 1988;\\
224, Langer et al., 1996;\\
225, Shepherd et al., 1997;\\
226, Magnier et al., 1999;\\                    %
227, Reipurth, 1999b;\\   
228, Cabrit \& Bertout, 1992; \\   
229, Bally et al., 1999;\\
230, Zhou et al., 1993;\\
231, Zhou, 1999;\\
232, Gregersen et al., 1997;\\
233, Walker et al., 1986;\\
234, Zhou, 1995;\\
235, Myers, et al., 1996;\\
236, Ward-Thompson et al., 1996; \\
237, Beuther et al., 2002a;\\
238, Arce \& Goodman, 2001;\\
239, Ridge \& Moore, 2001;\\
240, Yun \& Clemens, 1994;\\
241, Hirano et al., 1998;\\
242, Wolf-Chase et al., 2000;\\
243, Monin et al., 1996;\\
244, Palacios \& Eiroa, 1999;\\
245, Moriarty-Shieven \& Snell, 1988;\\
246, Hogerheijde et al., 1998\\
247, Hunter et al. 1994;\\
248, Bachiller et al., 1995a;\\
249, Chernin \& Wright, 1996;\\
250, Kawamura et al., 1999;\\
251, Reipurth, 1999; \\
252, Yamashita et al., 1989b;\\
253, Myers et al., 1995; \\%
254, Sugitani et al., 1991;\\
255, Churchwell, 1999;\\
256, Yamaguchi et al., 1999a;\\
257, Onishi et al., 1999;\\
258, Dobashi et al., 2001;\\
259, Yamaguchi et al., 1999b;\\
260, Saito et al., 1999;\\
261, Chernin \& Masson, 1995\\
262, Aso et al., 2000;\\
263, Moro-Mart\'\i n et al., 1999;\\
264, Davis et al., 2000;\\
265, Davis et al., 1994a;\\
266, Eisl\"offel, 2000;\\
267, Hodapp \& Ladd, 1995;\\
268, Davis et al., 1997;\\
269, Aspin, 1992;\\
270, Bachiller et al., 1994;\\
271, Hartigan et al., 1989;\\
272, Yu et al., 1999;\\
273, Stanke et al., 1998a;\\
274, Gomez et al. 1997;\\
275, Davis et al., 1994b;\\
276, Davis et al., 1995;\\
277, Yu et al., 1997;\\
278, Stanke et al., 2002;\\
279, Porras et al., 2000;\\
280, Davis et al., 2002;\\
281, Field et al., 1998;\\
282, Chen et al., 1999;\\
283, Luhman et al., 1997;\\
284, Chakraborty et al., 2000;\\
285, Massi et al., 1997;\\
286, Aspin, 1998;\\
287, Ryder et al., 1998;\\
288, Yun et al., 1997;\\
289, Heathcote et al., 1996;\\
290, Eisl\"offel et al., 2000;\\
291, Davis \& Eisl\"offel, 1995;\\
292, Casali et al., 1993;\\
293, Davis et al., 1999a;\\
294, Aspin et al., 1991;\\
295, Holbrook, 1996;\\
296, Stanke et al., 2000;\\
297, Garden et al., 1990;\\
298, Poetzel et al., 1992;\\
299, Ayala et al., 1998;\\
300, Nisini et al., 2001;\\
301, Yu et al., 1995;\\
302, Davis et al., 1998;\\
303, Stanke et al, 1998b;\\
304, Cabrit et al., 1997;\\
305, Hartigan et al., 1995;\\
306, Kumar et al., 2002;\\
307, Shepherd et al., 1998;\\
308, Bloomer et al., 1998;\\
309, Longmore et al., 1986;\\
310, Velusamy \& Langer, 1998;\\  %
311, Davis et al., 1996;\\
312, Mitchell \& Hasegawa, 1991; \\%
313, Hunter et al., 1995; \\
314, Beuther et al., 2002b\\
315, Qin \& Wu, 2003a; \\
316, Yu et al., 2000;\\
317, Fuente et al., 2001;\\
318, Bourke, 2001;\\
319, Arce \& Goodman, 2002;\\
320, Moreira et al., 2000;\\
321, Dobashi \& Uehara, 2001;\\
322, White et al., 1997;\\
323, Sandell \& Aspin, 1998;\\
324, Davis et al., 1999b;\\
325, Beuther et al., 2002c;\\
326, Aspin \& Reipurth, 2000;\\
327, Kelly \& Macdonald, 1996;\\
328, Ogura, Sugitani \& Pickles, 2002;\\
329, Zhou et al., 1996;\\
330, Ohashi et al., 1997;\\
331, Choi et al., 1999;\\
332, Bally et al., 2002;\\
333, Yang et al., 1997;\\
334, Gregersen et al., 2000;\\
335, Lada \& Fich, 1996;\\
336, Fich \& Lada, 1998;\\
337, Kurtz et al., 1994;\\
338, Walsh et al., 1998;\\
339, McCutcheon et al., 1991;\\
340, Altenhoff et al., 1978;\\
341, Gregory \& Condon, 1991;\\
342, Griffith et al., 1994;\\
343, Wright et al., 1994;\\
344, Wolf-Chase \& Gregersen, 1997;\\
345, Comoretto et al., 1990;\\
346, Brand et al., 1994;\\
347, Valdettaro et al., 2001;\\
348, Mitchell et al., 1991;\\
349, Braz \& Epchtein, 1983;\\
350, Scalise et al., 1989;\\
351, Bence et al., 1996;\\
352, Rodriguez \& Canto, 1982\\
353, Moran \& Rodr\'\i guez, 1980;\\
354, Mitchell, et al., 1995;\\
355, Hofner et al., 1996;\\
356, Mardones et al., 1997;\\
357, Olmi, Felli \& Cesaroni, 1997;\\
358, Bence et al., 1998;\\
359, Molinari et al, 1998;\\
360, Chandler \& Carlstrom, 1996;\\
361, Snell \& Bally, 1986;\\
362, Garden et al., 1991b;\\
363, Cesaroni et al., 1997;\\
364, Richer et al., 1992;\\
365, White et al., 1995;\\
366, Tamura et al, 1996;\\
367, Andr\'e et al., 1999;\\
368, Belloche et al., 2002;\\
369, Wootten et al., 1989.\\
370, Gueth et al., 2001;\\
371, Magnier et al., 1995;\\
372, Mao et al., 2002;\\
373, Beltr\'an et al., 2002;\\
374, Anglada \& Rodr\'\i guez, 2002;\\
375, Girart et al., 2002;\\
376, Zhao et al., 1999;\\
377, Hilton \& Lahulla, 1993;\\
378, Kamazaki et al., 2003;\\
379, Gibb \& Little, 2000;\\
380, Shi et al., 2003;\\
381, Yu et al., 2003a;\\
382, Zhao et al., 2003;\\
383, Lee et al., 2002;\\
384, Phillips \& Mampaso, 1996;\\
385, Dent \& Aspin, 1992;\\
386, Reipurth et al, 2000;\\
387, Bontemps et al., 1996b;\\
388, Chernin, 1996;\\
389, Lefloch, 1996;\\
390, Bally et al., 1996a;\\
391, Bally et al., 1996b;\\
392, Noriega et al., 2002;\\
393, Patel et al., 2000;\\
394, Khanzadyan et al., 2002;\\
395, Padman et al., 1997;\\
396, Siebenmorgen \& Kr\"ugel, 2000;\\
397, Yu et al., 2003b;\\
398, Clark, 1991;\\
399, Gueth et al., 1997;\\%
400, Bally et al, 1995;\\
401, Rodr\'\i guez \& Reipurth, 1996;\\
402, Momose et al., 1998;\\
403, Hilton \& Lahulla, 1995;\\
404, Chini et al., 1997;\\
405, McMullin et al., 1994;\\
406, Hurt \& Barsony, 1996;\\
407, McCaughrean et al., 1994;\\
408, Gibb \& Davis, 1998;\\
409, Bontemps et al., 1995;\\
410, Bachiller et al., 1995b;\\
411, Cernicharo \& Reipurth, 1996;\\
412, Nagar et al., 1997;\\
413, Devine et al., 1999;\\
414, Codella et al., 2001;\\
415, Narayanan et al., 2002;\\
416, Zinchenko et al., 1995.\\
417, Braz \& Scalise, 1982;\\
418, Furuya et al., 2003;\\
419, Sun \& Wu, 2003;\\
420, Zhu et al., 2003; \\
421, Qin \& Wu, 2003b;\\
422, Visser et al., 2001;\\
423, Visser et al., 2002;\\
424, Wu \& Evans, 2003.